\documentclass[12pt]{article}

\usepackage{newtxtext,newtxmath}
\usepackage{graphicx}

\usepackage[letterpaper,margin=1in]{geometry}

\renewenvironment{abstract}
	{\quotation}
	{\endquotation}

\date{}

\makeatletter
\renewcommand{\fnum@figure}{\textbf{Figure \thefigure}}
\renewcommand{\fnum@table}{\textbf{Table \thetable}}
\makeatother

\usepackage{scicite}

\usepackage{url}
\usepackage{tabularx}
\usepackage{booktabs}
\usepackage{multirow}
\usepackage{enumitem}
\usepackage{todonotes}
\newcommand{\be}{\begin{equation}}
\newcommand{\ee}{\end{equation}}
\newcommand{\bea}{\begin{eqnarray}}
\newcommand{\eea}{\end{eqnarray}}

\def\bc{\begin{center}}
\def\ec{\end{center}}
\newcommand{\avg}[1]{\langle{#1}\rangle}
\newcommand{\Avg}[1]{\left\langle{#1}\right\rangle}

\usepackage{xcolor}

\def\scititle{
Directionality and node heterogeneity reshape criticality in hypergraph percolation
}
\title{\bfseries \boldmath \scititle}

\author{
	Yunxue Sun$^{1}$,
	Xueming Liu$^{1\ast}$,
	Ginestra Bianconi$^{2\ast}$\and
	\small$^{1}$School of Artificial Intelligence and Automation, State Key Laboratory of Digital Manufacturing Equipment \and
    \small and Technology, Institute of Medical Equipment Science and Engineering,  \and
    \small Huazhong University of Science and Technology, Wuhan, China.\and
	\small$^{2}$School of Mathematical Sciences, Queen Mary University of London, London E1 4NS, UK.\and
	\small$^\ast$Corresponding author. Email: xm\_liu@hust.edu.cn; ginestra.bianconi@gmail.com
}

\begin{document} 

% Insert the title and author list
\maketitle

% Abstract, in bold
% There are strict length limits, and not all formats have abstracts.
% Consult the journal instructions to authors for details.
% Do not cite any references in the abstract.
\begin{abstract} \bfseries \boldmath
Directed and heterogeneous hypergraphs capture directional higher-order interactions with intrinsically asymmetric functional dependencies among nodes. As a result, damage to certain nodes can suppress entire hyperedges, whereas failure of others only weakens interactions. Metabolic reaction networks offer an intuitive example of such asymmetric dependencies.
Here we develop a message-passing and statistical mechanics framework for percolation in directed hypergraphs that explicitly incorporates directionality and node heterogeneity. Remarkably, we show that these hypergraph features have a fundamental effect on the critical properties of hypergraph percolation, reshaping criticality in a way that depends on network structure. Specifically, we derive anomalous critical exponents that depend on whether node or hyperedge percolation is considered in maximally correlated, heavy-tailed regimes.
These theoretical predictions are validated on synthetic hypergraph models and on a real directed metabolic network, opening new perspectives for the characterization of the robustness and resilience of real-world directed, heterogeneous higher-order networks.

\end{abstract}

\section*{INTRODUCTION}\label{sec1}

Percolation theory provides a foundational framework for understanding robustness, connectivity, and phase transitions in complex networks~\cite{Dorogovtsev2008Critical,Kahng2009Percolation,Li2021Percolation, Artime2024Robustness, Kryven2019Bond,Baxter2012Avalanche,watanabe2014cavity, PhysRevE.110.064303}. 
It characterizes how large-scale connectivity emerges, collapses, and reorganizes under structural perturbations~\cite{Buldyrev2010Catastrophic,Bianconi2018Multilayer, Meng2023Percolation,hu2025unveiling}, and has been widely used to study robustness and cascading phenomena in a broad range of systems~\cite{radicchi2015breaking, Cirigliano2024General,Kim2024Shortest,cirigliano2023extended,min2024no}. 
Recent advances have extended percolation theory beyond simple graphs to network representations that explicitly encode higher-order interactions.

Higher-order networks generalize graph-based representations by encoding interactions among more than two nodes, typically formalized as hypergraphs or simplicial complexes~\cite{Bianconi2021Higher,Battiston2020Networks,Bick2021Higher,Torres2021And, larockEncapsulationStructureDynamics2023, lambiotteNetworksOptimalHigherorder2019}. 
Such representations naturally arise in systems ranging from brain~\cite{Faskowitz2022Edges,santoro2023higher} and biochemical reaction networks~\cite{Jost2019Hypergraph} to protein–protein interactions~\cite{klimm2021hypergraphs}, social contagion~\cite{Iacopini2019Simplicial,landry2020effect,Ferraz2024Contagion}, and supply-chain dynamics~\cite{mungo2024reconstructing}. 
A growing body of work has shown that this collective organization fundamentally reshapes connectivity and robustness~\cite{Sun2021Higher,Bianconi2024Nature,Sun2023Dynamic,lee2023k,2025percolation}, leading to percolation behavior—including discontinuous, hybrid, and multi-stage transitions—that has no analog in simple graphs.

Despite these advances, most higher-order percolation models assume undirected interactions. 
In contrast, many real-world systems are inherently directional: chemical reactions proceed from reactants to products and regulatory cascades transmit information in a prescribed order. Introducing directionality into percolation theory has already transformed our understanding of pairwise networks~\cite{Newman2001Random,dorogovtsev2001giant}, revealing distinct Giant In (GIN), Giant Out (GOUT), and Giant Strongly Connected (GSCC) components~\cite{liu2017controllability}, as well as asymmetric thresholds, tricritical and quadruple points, and multi-phase transitions~\cite{azimi2014giant,Liu2016Breakdown,liu2019multiple,liu2023possible,liu2020robustness}. However, understanding how directionality affects the robustness of higher-order systems is therefore a key open challenge.

Beyond directionality, many higher-order systems feature heterogeneous node roles, where the failure of certain indispensable participants suppresses an entire interaction rather than merely reducing its cardinality~\cite{Bianconi2024Nature,2025percolation, sun2024higher}. In our framework, we model these indispensable participants as anchor nodes.
They capture situations where, for example, a chemical reaction cannot proceed without its catalyst or a production process halts when a critical component fails. 
From a percolation perspective, anchor nodes represent intrinsic points of fragility that govern both the percolation threshold and the critical behavior of higher-order networks~\cite{radicchi2015breaking}.

In this work, we develop a unified theoretical framework for percolation on directed hypergraphs with anchor nodes. 
Using a message-passing formalism~\cite{Bianconi2018Multilayer,Karrer2010Message,Newman2023Message, ruggeriMessagepassingHypergraphsDetectability2024}, we derive analytical conditions for the emergence of the Hypergraph Giant In Component (HGIN), Hypergraph Giant Out Component (HGOUT), and Hypergraph Giant Strongly Connected Component (HGSCC). 
We show that anchor nodes fundamentally alter the percolation properties by separating node- and hyperedge-level thresholds, which are not distinguished in their absence. 
As the prevalence of anchor nodes increases, the system becomes increasingly fragile, with the node percolation threshold shifting to higher values.
While the percolation transition remains continuous, directionality and functional dependencies reorganize critical behavior across different hypergraph components. 
For well-behaved degree and cardinality distributions, we identify a composition rule for the critical exponents, whereby the exponent of the HGSCC follows an additive relation involving the corresponding In- and Out-Components,
$\beta_{\mathrm{HGSCC}} = \beta_{HGOUT} + \beta_{HGIN}$. 
We further show that this relation is modified in maximally correlated topologies with heavy-tailed distributions. 
Together, our results clarify how directionality and functional dependencies jointly shape universality and robustness in higher-order network percolation.

\section*{RESULTS}\label{sec2}

\begin{figure*}
    \centering
    \includegraphics[width=1\linewidth]{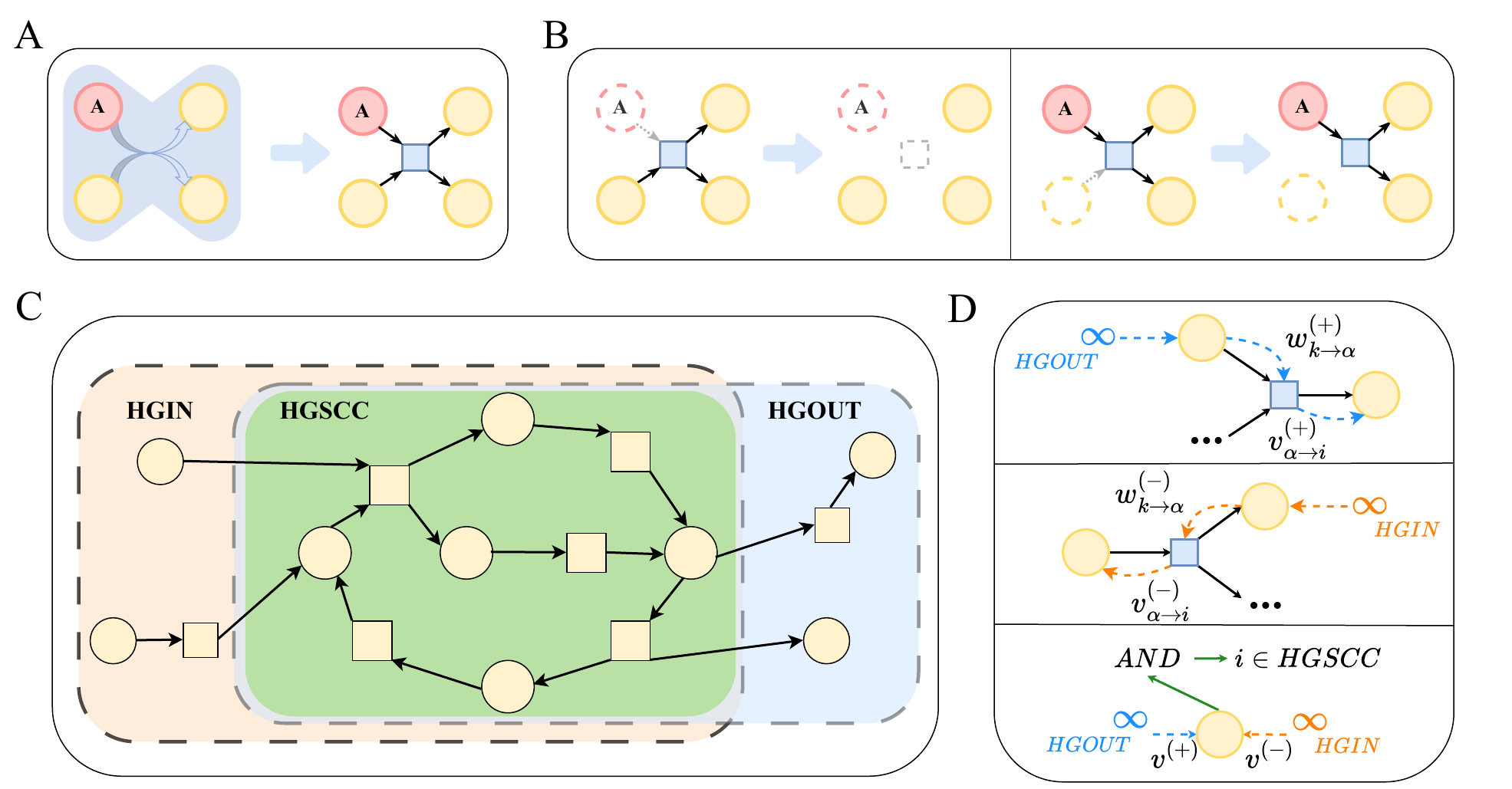} % 
    \caption{
\textbf{Directionality and functional dependencies in hypergraph percolation.}
\textbf{(A)} Representation of a directed hyperedge as a bipartite factor graph, where directionality encodes collective interactions from input nodes to output nodes. Anchor nodes (red) act as indispensable elements that control hyperedge activation, whereas non-anchor nodes (yellow) participate without determining functionality.
\textbf{(B)} Anchor-node--induced functional dependency. Removal of an anchor node deactivates the entire hyperedge, while removal of a non-anchor node only weakens the interaction by reducing hyperedge cardinality.
\textbf{(C)} Macroscopic organization of connectivity in a directed hypergraph, illustrating the three giant components: the HGSCC, comprising mutually reachable nodes and hyperedges; the HGIN, consisting of elements that can reach the HGSCC; and the HGOUT, containing elements reachable from the HGSCC.
\textbf{(D)} Directional message-passing processes underlying percolation. Forward propagation ($w^{(+)}, v^{(+)}$) identifies the HGOUT, backward propagation ($w^{(-)}, v^{(-)}$) identifies the HGIN, and nodes reached by both processes form the HGSCC.
}
    \label{fig:schematic}
\end{figure*}

%\subsection*{Message-passing theory for percolation on directed hypergraphs}
\subsection*{Percolation in directed hypergraphs with functional dependencies}

We study percolation in directed hypergraphs, a class of higher-order networks in which interactions involve multiple nodes and propagate along prescribed directions.
A directed hypergraph comprises $N$ nodes and $M$ directed hyperedges, each connecting a set of input nodes to a set of output nodes and thereby defining an asymmetric flow of influence.
Such systems are naturally characterized by a joint node degree distribution $P(q^{\rm in},q^{\rm out})$ and a hyperedge cardinality distribution $Q(m^{\rm in},m^{\rm out})$.
Directed hypergraphs provide a natural framework for modeling systems in which functionality emerges from collective, directional interactions, including biochemical reaction networks, ecological processes, and information-processing systems.

To analyze connectivity and robustness, we map the directed hypergraph onto a directed bipartite factor graph (Fig.~\ref{fig:schematic}\textbf{A}), in which nodes and hyperedges form distinct vertex classes connected by directed links. This representation preserves the interaction directionality while yielding a locally tree-like architecture that enables an analytical message-passing treatment.

Beyond topological directionality, many real systems feature interactions whose functionality depends asymmetrically on participating components rather than on connectivity alone. We capture this functional asymmetry by introducing anchor nodes (Fig.~\ref{fig:schematic}\textbf{B}), defined as components whose removal suppresses an entire interaction, whereas removal of other nodes only reduces its effective cardinality. Within this framework, each node associated with a hyperedge is designated as an anchor independently with probability $\theta$, providing a simple statistical encoding of role-based functional constraints at the level of higher-order interactions.

We probe robustness under random damage by allowing nodes and hyperedges to survive independently with probabilities $p_N$ and $p_H^{[\mathbf{m}_{\alpha}]}$, respectively.
As order parameters of the percolation transition, we consider the relative sizes of the HGIN, HGOUT, and HGSCC (Fig.~\ref{fig:schematic}\textbf{C}).
These quantities generalize the notion of directed giant components in pairwise networks to higher-order systems with functional dependencies.

Connectivity is characterized analytically through a message-passing framework defined on the directed factor graph (Fig.~\ref{fig:schematic}\textbf{D}).
The dynamics decompose naturally into two directional propagation processes.
Forward messages propagate along the interaction direction and determine reachability within the HGOUT, whereas backward messages propagate against the interaction direction and determine the HGIN.
The corresponding local update rules, incorporating both directionality and anchor-induced functional dependencies, read
\begin{equation}
\label{m_anchor_e}
\begin{aligned}
w_{i\rightarrow \alpha}^{(+)} &= p_N^{1-z_{i\alpha}}
\left[ 1 - \prod_{\beta\in N^{(-)}(i)\setminus \alpha}
(1 - v_{\beta\rightarrow i}^{(+)}) \right], \\
v_{\alpha\rightarrow i}^{(+)} &= p_H^{[\mathbf{m}_{\alpha}]} p_N^{Z_{i\alpha}}
\left[ 1 - \prod_{j\in N^{(-)}(\alpha)\setminus i}
(1 - w_{j\rightarrow \alpha}^{(+)}) \right], \\
w_{i\rightarrow \alpha}^{(-)} &= p_N^{1-z_{i\alpha}}
\left[ 1 - \prod_{\beta\in N^{(+)}(i)\setminus \alpha}
(1 - v_{\beta\rightarrow i}^{(-)}) \right], \\
v_{\alpha\rightarrow i}^{(-)} &= p_H^{[\mathbf{m}_{\alpha}]} p_N^{Z_{i\alpha}}
\left[ 1 - \prod_{j\in N^{(+)}(\alpha)\setminus i}
(1 - w_{j\rightarrow \alpha}^{(-)}) \right],
\end{aligned}
\end{equation}
where $z_{i\alpha}$ indicates whether node $i$ acts as an anchor for hyperedge $\alpha$, and $Z_{i\alpha}$ denotes the number of other anchor nodes within the same hyperedge.
Full derivations are provided in section \ref{ApA}.

At the fixed point of these equations, nodes and hyperedges that are simultaneously reachable by both forward and backward propagation form the HGSCC.
Importantly, membership in the HGSCC does not factorize into independent contributions from the HGIN and HGOUT, because node survival is a shared prerequisite for both propagation processes.
Consequently, the probabilities that a node or hyperedge belongs to the HGSCC are given by
\begin{equation}
\begin{split}
r_i &= p_N
\left[ 1 - \prod_{\alpha\in N^{(+)}(i)}
(1 - v_{\alpha \rightarrow i}^{(-)}) \right]
\left[ 1 - \prod_{\alpha\in N^{(-)}(i)}
(1 - v_{\alpha \rightarrow i}^{(+)}) \right], \\
s_{\alpha} &= p_H^{[\mathbf{m}_{\alpha}]} p_N^{Z_{\alpha}}
\left[ 1 - \prod_{i\in N^{(-)}(\alpha)}
(1 - w_{i\rightarrow \alpha}^{(+)}) \right]
\left[ 1 - \prod_{i\in N^{(+)}(\alpha)}
(1 - w_{i\rightarrow \alpha}^{(-)}) \right].
\end{split}
\end{equation}
Averaging these quantities over nodes and hyperedges yields the macroscopic sizes of the HGIN, HGOUT, and HGSCC, which serve as global indicators of the percolation transition.

\begin{figure}
	\centering % 
    \includegraphics[width=1\linewidth]{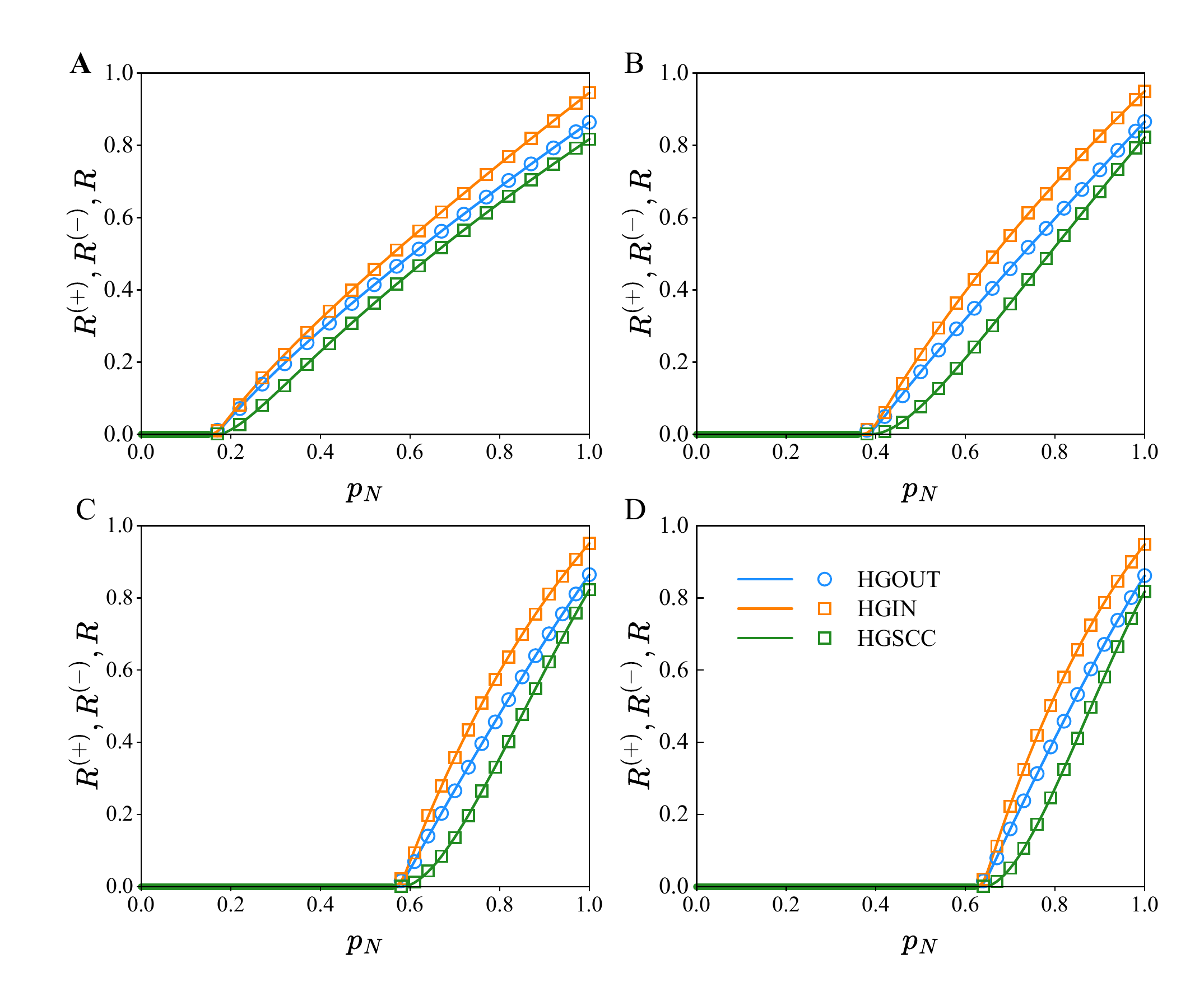} % 
	\caption{\textbf{Percolation in directed hypergraphs with increasing anchor-node fraction.}
Relative sizes of the HGOUT, HGIN, and HGSCC are shown as functions of the node survival probability $p_N$ for synthetic directed hypergraphs with anchor-node probability $\theta$.
Symbols denote Monte Carlo simulations, while solid lines indicate analytical predictions from the message-passing theory.
Increasing $\theta$ systematically shifts the percolation threshold toward larger $p_N$, indicating enhanced structural fragility and reduced robustness.
Panels correspond to $\theta = 0$ (A), $0.4$ (B), $0.8$ (C), and $1$ (D).
All results are obtained for random directed hypergraphs with $N = 10^4$ nodes and $M = 10^4$ hyperedges, with fixed cardinalities $m^{\mathrm{\rm in}} = 3$ and $m^{\mathrm{\rm out}} = 2$.
	}
    \label{fig:r1} 
\end{figure}

We compare the predictions of the message-passing theory with numerical results obtained for synthetic directed hypergraphs.
To isolate the effect of functional dependencies, we set $p_H^{[\mathbf{m}_{\alpha}]}=1$ and focus on node percolation.
Figure~\ref{fig:r1} presents the relative sizes of the HGIN, HGOUT, and HGSCC as functions of the node survival probability $p_N$ for representative values of the anchor-node fraction.

The analytical predictions are in close agreement with numerical simulations, validating the message-passing framework. Standard formulations of message passing rely on the assumption that neighbors' states are uncorrelated, a condition that breaks down in networks with loops \cite{cantwell2019message}. Remarkably, despite the extensive loops inherent to the HGSCC, our framework accurately captures its emergence and scaling behavior, demonstrating a robustness that transcends the standard tree-like assumption.
As the anchor-node fraction increases, the emergence of all three giant components is systematically shifted toward larger values of $p_N$, indicating that anchor-induced functional dependencies hinder the formation of macroscopic connectivity.
In the following section, we show how this mechanism quantitatively reshapes the percolation threshold and the associated critical behavior.

\subsection*{Percolation thresholds shaped by directionality and node heterogeneity}

Macroscopic connectivity in directed hypergraphs emerges through a linear instability of the message-passing dynamics.
At the percolation threshold, the trivial fixed point associated with a fully fragmented state becomes unstable, marking the onset of giant connected components.
Although information and influence propagate asymmetrically at the microscopic level, the forward and backward propagation channels are governed by non-backtracking operators with an identical leading eigenvalue (see section \ref{sec:S3}).
As a result, the HGIN, HGOUT, and HGSCC do not arise separately but appear simultaneously at a single critical point, when the effective branching factor $\hat{\Lambda}$ exceeds unity,
\begin{equation}
\begin{split} 
  \hat{\Lambda}
  = p_N\frac{\avg{q^{\rm in}q^{\rm out}}}{\avg{q^{\rm out}}}
    \frac{\avg{{p}_H^{[\bf m]}{\pi}_N^{m-2}m^{\rm out}m^{\rm in}}}{\avg{m^{\rm out}}}
  > 1,
\end{split}
\label{thre}
\end{equation}
where $p_N$ and $p_H^{[\mathbf{m}]}$ denote the survival probabilities of nodes and hyperedges, respectively, and 
$\pi_N = 1 - \theta + \theta p_N$ is the probability that a node is either a non-anchor or an intact anchor.
It makes explicit how interaction directionality, higher-order connectivity, and anchor-induced functional dependencies collectively determine the percolation threshold. The detailed derivation of this criterion are provided in section \ref{si-aver}.

Equation~\ref{thre} clarifies how anchor nodes modulate global connectivity.
The factor $\pi_N^{m-2}$ introduces a selective suppression of large hyperedges, an effect that strengthens monotonically with increasing anchor probability $\theta$.
Consequently, maintaining macroscopic connectivity requires a progressively larger fraction of surviving nodes, yielding a systematic upward shift of the percolation threshold in $p_N$.
This behavior reflects the inherent fragility of conditional interactions, in which the failure of a single indispensable component can disable an entire higher-order interaction.
By contrast, in anchor-free systems—where node and hyperedge percolation thresholds coincide—anchor-induced functional dependencies break this symmetry and selectively elevate the node percolation threshold.

\begin{figure}
	\centering % 
    \includegraphics[width=1\linewidth]{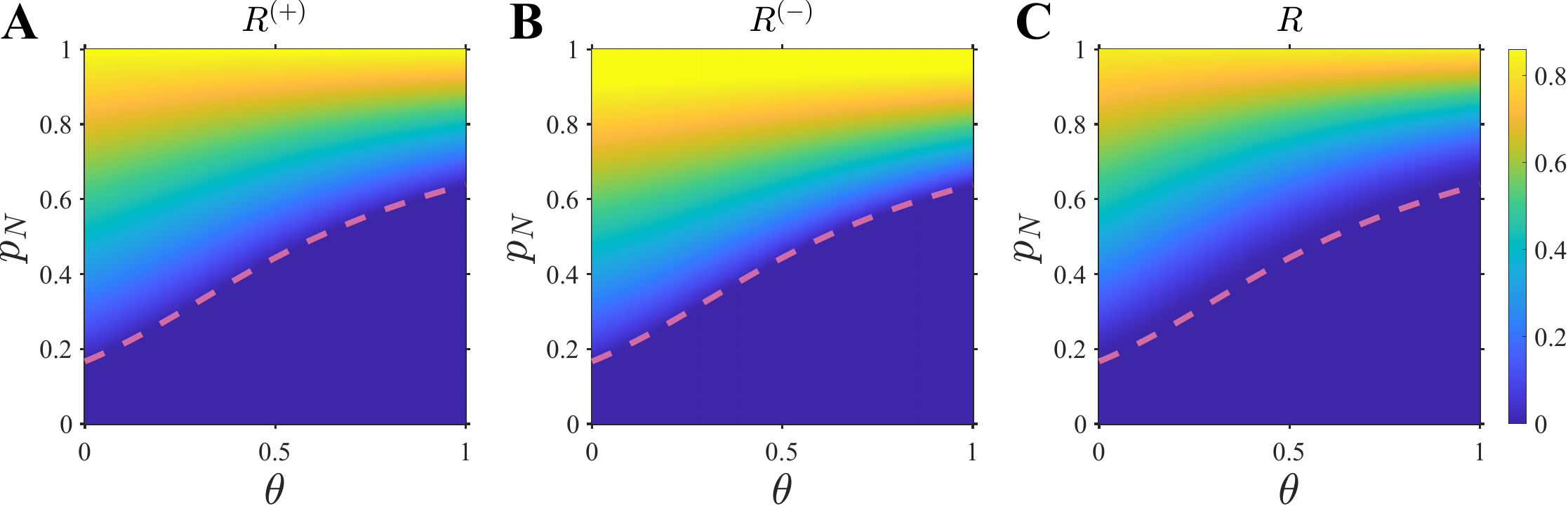} % 
\caption{
\textbf{Phase diagrams of directed hypergraph percolation with anchor-node heterogeneity.}
Equilibrium sizes of the \textbf{(A)} HGOUT ($R^{(+)}$), \textbf{(B)} HGIN ($R^{(-)}$), and \textbf{(C)} HGSCC ($R$) are shown over the parameter space defined by the node survival probability $p_N$ and the anchor-node probability $\theta$.
Colors indicate the relative size of the corresponding giant component, ranging from the fragmented phase ($R \approx 0$) to the percolating phase.
Dashed lines denote the theoretical percolation thresholds obtained from linear stability analysis.
All results are shown for random directed hypergraphs with $N = 10^4$ nodes and $M = 10^4$ hyperedges, with fixed cardinalities $m^{\mathrm{\rm in}} = 3$ and $m^{\mathrm{\rm out}} = 2$.
Although all three giant components emerge at the same critical boundary, their post-critical growth is governed by distinct critical exponents, with the HGSCC exhibiting systematically slower scaling.
}
	\label{fig:heatmap} 
\end{figure}

These effects are summarized in Fig.~\ref{fig:heatmap}, which displays the phase diagrams of the HGOUT, HGIN, and HGSCC in the $(p_N,\theta)$ plane.
The critical line predicted by Eq.~\ref{thre} accurately separates the fragmented and percolating phases.
As $\theta$ increases, the connected region contracts, indicating that anchor nodes systematically erode global connectivity.
Although all three giant components emerge at the same critical boundary, their post-critical behavior differs substantially: the HGIN and HGOUT grow rapidly above threshold, whereas the HGSCC develops more gradually.
This contrast anticipates distinct critical exponents governing the different components, which we analyze in the following section.

\subsection*{Anomalous critical scaling and universality classes}

Beyond the location of the percolation threshold, the nature of the transition is encoded in the critical scaling of the emerging giant components.
In directed hypergraphs, percolation gives rise to three macroscopic structures which appear simultaneously at criticality but display markedly different growth behaviors above the transition.
This separation directly reflects the intrinsic asymmetry of directed higher-order interactions.

Close to the critical point, $\hat{\Lambda}\gtrsim 1$, the sizes of the HGIN and HGOUT exhibit power-law scaling,
\begin{equation}
R^{(\pm)} \propto (\hat{\Lambda}-1)^{\beta^{(\pm)}},
\end{equation}
where $\beta^{(+)}$ and $\beta^{(-)}$ characterize forward and backward propagation, respectively.
By contrast, the HGSCC arises from the intersection of these two directional structures and is governed by a distinct exponent $\beta$,
\begin{equation}
R \propto (\hat{\Lambda}-1)^{\beta}.
\end{equation}
Detailed derivations are provided in Section~\ref{Apc}.

A central result is that these exponents are not independent.
For a broad class of directed hypergraphs—including uncorrelated networks and correlated systems with finite moments—the critical exponent of the HGSCC obeys a simple additive law,
\begin{equation}
\beta = \beta^{(+)} + \beta^{(-)}.
\label{eq:additive_law}
\end{equation}
This composition reflects the directed nature of connectivity: near criticality, belonging to the HGSCC requires simultaneously reaching and being reachable from the macroscopic structure.
These two conditions correspond to statistically independent branching processes, whose intersection leads to additive scaling.
In this regime, the system exhibits mean-field behavior, with $\beta^{(+)}=\beta^{(-)}=1$ and hence $\beta=2$, explaining the notably slower post-critical growth of the HGSCC compared with its directional counterparts.

\begin{figure}
    \centering
    \includegraphics[width=0.8\linewidth]{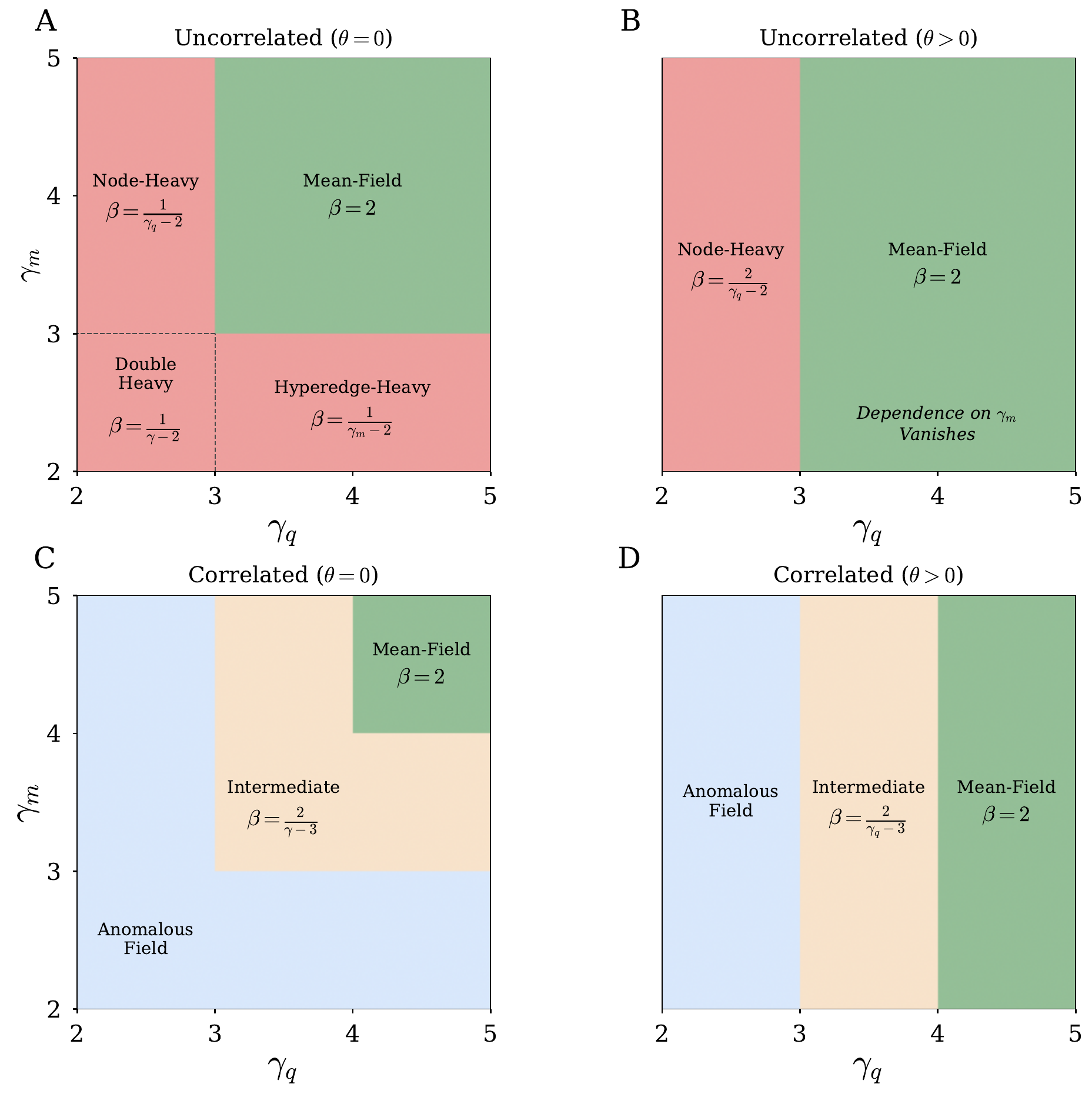}
\caption{
\textbf{Universality map of directed hypergraph percolation.}
Universality classes of the critical exponent $\beta$ in the parameter space defined by the node degree exponent $\gamma_q$ and the hyperedge cardinality exponent $\gamma_m$.
Panels correspond to \textbf{(A)} uncorrelated networks without anchor nodes ($\theta = 0$), 
\textbf{(B)} uncorrelated networks with anchor nodes ($\theta > 0$), 
\textbf{(C)} maximally correlated networks without anchor nodes ($\theta = 0$), and 
\textbf{(D)} maximally correlated networks with anchor nodes ($\theta > 0$).
Colored regions delineate distinct universality classes, including mean-field, node-dominated, hyperedge-dominated, intermediate, and anomalous regimes.
Analytical expressions for $\beta$ are indicated within each region.
}
    \label{fig:psp}
\end{figure}

The scaling behavior changes qualitatively in structurally heterogeneous networks.
We focus on scale-free hypergraphs, characterized by power-law in-degree (out-degree) and cardinality distributions
$P(q)\sim q^{-\gamma_q}$ and $Q(m)\sim m^{-\gamma_m}$.
In the presence of anchor nodes ($\theta>0$), functional constraints suppress the effect of large hyperedges, and the effective scaling behavior is governed by the node-degree exponent $\gamma=\gamma_q$.
In contrast, in the anchor-free limit ($\theta=0$), structural heterogeneity is controlled by the smaller of the two exponents, $\gamma=\min(\gamma_q,\gamma_m)$.
These distinctions define the relevant universality classes summarized in Fig.~\ref{fig:psp}.

For uncorrelated or weakly correlated networks with $\gamma>3$, the additive law~\eqref{eq:additive_law} remains valid.
However, in maximally correlated scale-free networks with $\gamma<3$, the order parameters acquire explicit dependence on the control variables, signaling a breakdown of standard universality.
In node percolation, this leads to an anomalous correction of the form
\begin{equation}
\beta_R = \beta_R^{(+)} + \beta_R^{(-)} - 1,
\end{equation}
whereas the hyperedge order parameter retains additive scaling.
In hyperedge percolation, the roles of nodes and hyperedges are reversed.
These anomalous regimes, which highlight the distinct dynamical roles of nodes and hyperedges in directed higher-order systems, are summarized in Tables~\ref{tab:finite_threshold} and~\ref{tab:zero_threshold}.

\begin{figure}
	\centering % 
    \includegraphics[width=0.8\linewidth]{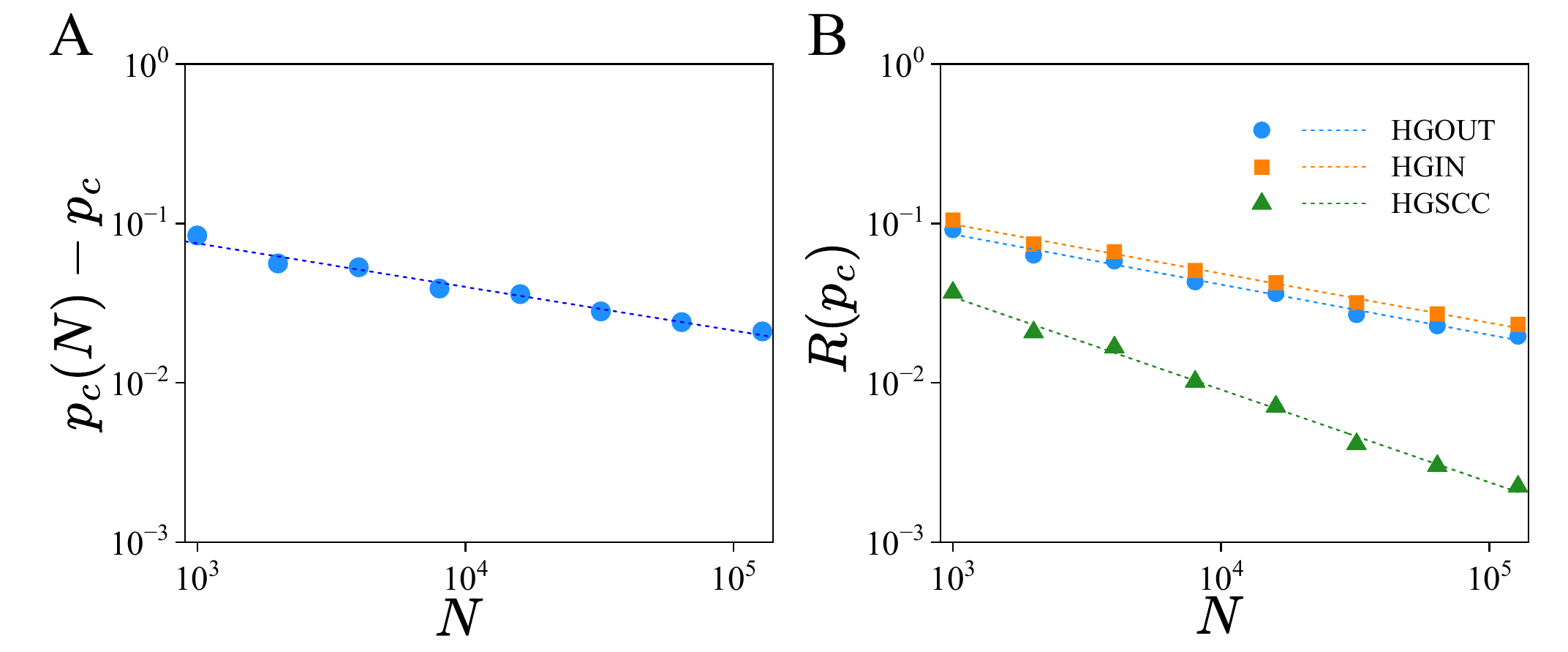} % 
\caption{
\textbf{Finite-size scaling in directed hypergraph percolation.}
\textbf{(A)} Scaling of the pseudo-critical point, showing the offset $p_c(N) - p_c$ as a function of network size $N$.
\textbf{(B)} Scaling of the percolation strength $R(p_c)$, evaluated at the pseudo-critical point, as a function of $N$.
Symbols show numerical results for the HGOUT (circles), HGIN (squares), and HGSCC (triangles).
Dashed lines indicate power-law fits, demonstrating critical scaling behavior.
All results are obtained for random directed hypergraphs with parameters identical to those in Fig.~~\ref{fig:r1}.
}
	\label{fig:FSS} 
\end{figure}

Finite-size scaling analyses~\cite{Blanc1986} further support these theoretical predictions.
Guided by the theoretical prediction of simultaneous emergence, we employ a unified scaling strategy in which the HGIN, HOUT, and HGSCC are evaluated at a common size-dependent pseudo-critical point (see Methods). As shown in Fig.~\ref{fig:FSS}, their post-critical growth is governed by distinct exponents.
In particular, the larger exponent associated with the HGSCC explains its visibly more gradual emergence compared with the rapid growth of the directional components.
Additional validation results for other topological regimes are reported in Section~\ref{ce}.

\begin{table}
\centering
\caption{Critical exponents for uncorrelated distributions with finite moments and correlated distributions with $\gamma > 3$. The order parameters scale as $R \propto (\hat{\Lambda}-1)^{\beta}$}
\label{tab:finite_threshold}
\begin{tabularx}{\textwidth}{@{\extracolsep{\fill}}llcc@{}} 
\toprule
\textbf{Network Topology} & \textbf{$\gamma$} & \textbf{$\beta^{(\pm)}$} & \textbf{$\beta$} \\
\midrule
1. Distributions with Finite Moments & All moments finite & $1$ & $2$ \\
\midrule
\multirow{2}{*}{2. Uncorrelated Power-law} & $\gamma > 3$ & $1$ & $2$ \\
& $\gamma \in (2, 3)$ & $1/(\gamma-2)$ & $2/(\gamma-2)$ \\
\midrule
\multirow{2}{*}{3. Maximally Correlated Power-law} & $\gamma > 4$ & $1$ & $2$ \\
& $\gamma \in (3, 4)$ & $1/(\gamma-3)$ & $2/(\gamma-3)$ \\
\bottomrule
\addlinespace
\multicolumn{4}{l}{\small \textbf{Note:} The exponent $\gamma$ is defined as $\gamma = \gamma_q$ if $\theta > 0$, and $\gamma = \min(\gamma_q, \gamma_m)$ if $\theta = 0$.} \\
\end{tabularx} 
\end{table}

\begin{table}
\centering
\caption{Critical exponents for maximally correlated distributions with $\gamma < 3$.}
\label{tab:zero_threshold}

\resizebox{\textwidth}{!}{
\begin{tabular}{@{}lccccccc@{}}
\toprule
\textbf{Condition} & $\beta_W^{(\pm)}$ & $\beta_V^{(\pm)}$ & $\beta_R^{(\pm)}$ & $\beta_S^{(\pm)}$ & $\beta_R$  & $\beta_S$ \\
\midrule

\multicolumn{7}{l}{\textit{\textbf{Node Percolation ($p_H=1$)}}} \\
\addlinespace
Cond. A & {$(\gamma - 2)/(3-\gamma)$} & {$(\gamma - 2)/(3-\gamma)$} & $1 + \beta_W^{(+)}$ & $\beta_W^{(+)}$ & $1 + 2\beta_W^{(+)}$ & $2\beta_W^{(+)}$ \\
Cond. B & $\{[(\gamma_q-2)(\gamma_m-2)]^{-1}-1\}^{-1}$ & $(\gamma_m-2)(1+\beta_W^{(\pm)})$ & $1 + \beta_W^{(+)}$ & $\beta_V^{(+)}$ & $1 + 2\beta_W^{(+)}$ & $2\beta_V^{(+)}$ \\
\midrule

\multicolumn{7}{l}{\textit{\textbf{Hyperedge Percolation ($p_N=1$)}}} \\
\addlinespace 
Cond. A & $1/(3-\gamma)$ & $1/(3-\gamma)$ & $\beta_W^{(+)}$ & $1 + \beta_W^{(+)}$ & $2\beta_W^{(+)}$ & $1 + 2\beta_W^{(+)}$ \\
Cond. B & $(\gamma_q-2)\beta_V^{(\pm)}$ & $[1-(\gamma_q-2)(\gamma_m-2)]^{-1}$ & $\beta_W^{(+)}$ & $1 + \beta_V^{(+)}$ & $2\beta_W^{(+)}$ & $1 + 2\beta_V^{(+)}$ \\
\bottomrule

\addlinespace
\multicolumn{7}{l}{\small \textbf{Note:} Conditions are defined based on the anchor probability $\theta$ and power-law exponents $\gamma_q, \gamma_m$:} \\
\multicolumn{7}{l}{\small \textbf{Cond. A:} $\gamma_q < 3$ and $\theta > 0$; OR $\theta=0$ and ( ($\gamma=\gamma_m < 3$ and $\gamma_q > 3$) OR ($\gamma=\gamma_q < 3$ and $\gamma_m > 3$) ).} \\
\multicolumn{7}{l}{\small \textbf{Cond. B:} $\theta=0$ and $\gamma_q < 3$ and $\gamma_m < 3$.} \\
\end{tabular}
} 
\end{table}

\subsection*{Validation in real directed hypergraphs}

\begin{figure}
    \centering
    \includegraphics[width=1\linewidth]{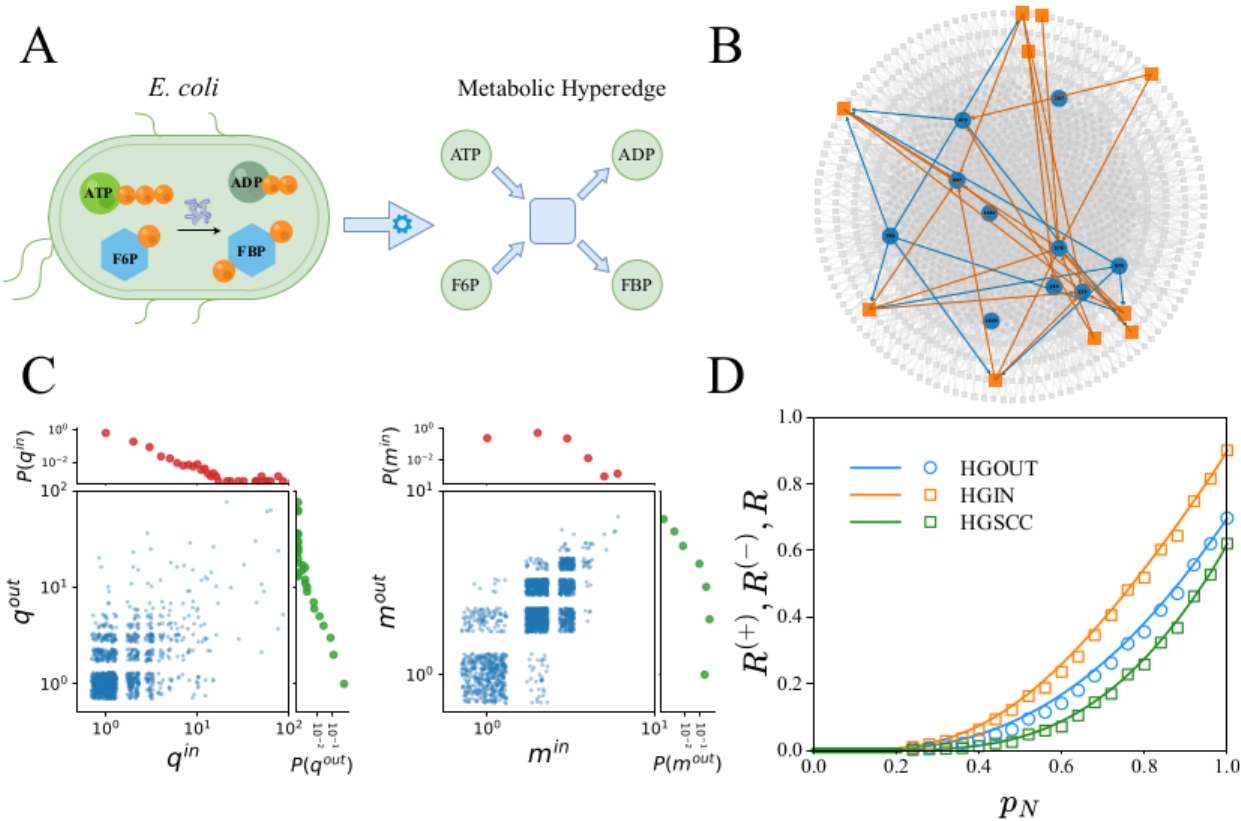}
\caption{
\textbf{Percolation in the \textit{E. coli} metabolic directed hypergraph.}
\textbf{(A)} Schematic illustration of the metabolic hypergraph representation, where biochemical reactions are modeled as directed hyperedges mapping substrate metabolites to product metabolites.
\textbf{(B)} Structural organization of the \textit{E. coli} metabolic hypergraph. The gray background shows the full network, while the colored foreground highlights a representative local subgraph in the factor-graph representation, with circles denoting metabolites and squares denoting reactions.
\textbf{(C)} Empirical degree and cardinality correlations in the metabolic hypergraph. Scatter plots show the joint distributions of node in- and out-degrees ($q^{\mathrm{\rm in}}$ vs. $q^{\mathrm{\rm out}}$, left) and hyperedge input and output cardinalities ($m^{\mathrm{\rm in}}$ vs. $m^{\mathrm{\rm out}}$, right). Marginal distributions are shown in the adjoining panels.
\textbf{(D)} Percolation behavior of the metabolic hypergraph under node removal. Relative sizes of the HGOUT, HGIN, and HGSCC are shown as functions of the node survival probability $p_N$, comparing analytical predictions (lines) with numerical results (symbols).
}
    \label{fig:ijo}
\end{figure}

Biologically constrained systems provide a stringent testbed for theories of directed higher-order percolation.
We therefore apply our framework to the \textit{E. coli} metabolic network~\cite{zhangLinkPredictionPredicting2018}, a canonical example in which interactions are governed by strict stoichiometric rules.
The network comprises $N=1805$ metabolites and $M=2253$ biochemical reactions and is naturally represented as a directed hypergraph, where hyperedges encode reactions and directionality follows the thermodynamic flow from substrates to products (Fig.~\ref{fig:ijo}\textbf{A, B}).

Unlike the random hypergraph ensembles considered above, the metabolic network exhibits a highly organized architecture, featuring clustering, degree heterogeneity, and strong correlations between inputs and outputs (Fig.~\ref{fig:ijo}\textbf{C}).
Such structural features are well known to challenge mean-field and locally tree-like approaches.
Here, biological functionality is incorporated directly into the hypergraph description through a physically grounded definition of anchor nodes.
Stoichiometry dictates that substrates are indispensable: the failure of any required substrate completely suppresses the associated reaction.
By identifying substrate metabolites as anchors, these functional dependencies are explicitly encoded in the network structure, yielding a high effective anchor density $\theta \approx 0.48$.

Despite the presence of strong correlations and extensive functional constraints, the analytical predictions remain in close agreement with numerical results across the full range of node survival probabilities $p_N$ (Fig.~\ref{fig:ijo}\textbf{D}).
The theory accurately captures the emergence of the HGIN, HGOUT, and HGSCC, demonstrating that the message-passing framework remains predictive even in a realistic, highly structured biological network.

Topological analysis reveals a pronounced asymmetry between node- and hyperedge-level statistics (Fig.~\ref{fig:ijo}\textbf{C}).
Node in- and out-degrees follow power-law distributions with exponents $\gamma_q^{\rm in}\approx 3.20$ and $\gamma_q^{\rm out}\approx 3.75$, ensuring finite second moments.
By contrast, hyperedge input and output cardinalities display much heavier tails, with exponents $\gamma_m^{\rm in}\approx 2.57$ and $\gamma_m^{\rm out}\approx 2.37$.
From a purely structural perspective, such heavy-tailed hyperedges would place the system in an anomalous scaling regime.
Indeed, in an anchor-free null model ($\theta=0$), the divergence of hyperedge moments violates the additive scaling law, leading to the corrected exponent $\beta=\beta^{(+)}+\beta^{(-)}-1$ (Condition~A, Table~2).

The empirical metabolic network, however, operates in a qualitatively different regime.
The high density of anchor nodes suppresses the influence of large hyperedges by enforcing strict functional dependencies at the reaction level.
As a result, contributions from heavy-tailed cardinalities are exponentially attenuated, effectively regularizing the percolation process.
This mechanism restores the standard universality class, in which the critical exponent obeys the additive relation $\beta=\beta^{(+)}+\beta^{(-)}$ without anomalous corrections.
Biological constraints therefore do more than shift critical thresholds: they actively reshape the universality of the transition.

This regularization comes at the cost of pronounced structural fragility.
Comparison with an anchor-free null model (Fig.~\ref{fig:nonanchor}) reveals a substantial upward shift of the percolation threshold, indicating that global connectivity is sustained only when a large fraction of metabolites remains intact.
Functionally, this reflects an intrinsic vulnerability of metabolic organization: because reactions depend on specific substrates rather than interchangeable inputs, the loss of a small subset of metabolites can trigger a disproportionate collapse of metabolic capability.

At the same time, anchor-induced dependencies ensure a controlled mode of functional recovery.
Unlike the abrupt, explosive transition predicted for the anomalous null model, the empirical network exhibits a smooth and continuous growth of the giant components above criticality.
Once the critical nutrient availability is reached, metabolic functionality is restored progressively rather than catastrophically, highlighting a fundamental trade-off between robustness and controllability in biologically constrained systems.

\section*{DISCUSSION}\label{sec12}

This work demonstrates that criticality in higher-order systems is governed not only by interaction density, but by the logical structure through which interactions are functionally realized.
In directed hypergraphs, connectivity is intrinsically asymmetric: nodes contribute to collective organization by receiving influence, transmitting it, or participating in both processes.
When this directionality is coupled with functional constraints encoded by anchor nodes, the emergence of global connectivity departs qualitatively from that of undirected or purely structural networks.
As a consequence, robustness in directed higher-order systems cannot be understood without explicitly accounting for both the direction of influence and the functional indispensability of individual components.

Anchor nodes fundamentally reshape failure mechanisms in higher-order interactions.
Rather than depending on the presence of any generic subset of participants, interactions become conditional on the simultaneous availability of specific nodes.
The failure of a single indispensable element can therefore deactivate an entire interaction, even when the surrounding network remains largely intact.
This logic breaks the symmetry between node and hyperedge percolation that characterizes anchor-free systems, leading to a decoupling of their critical thresholds.
As a result, networks with dense connectivity may nonetheless exhibit pronounced fragility when functionality hinges on a critical fraction of anchored components.

Beyond shifts in the percolation threshold, the critical exponents of directed hypergraph percolation reveal how structural heterogeneity, interaction directionality, and functional constraints jointly shape macroscopic connectivity.
Unlike undirected hypergraphs—where node- and hyperedge-based order parameters typically share the same mean-field scaling—directed systems support distinct critical behaviors associated with forward and backward propagation.
Anchor nodes further modify this landscape by tying the activation of higher-order interactions to specific nodes, thereby coupling structural heterogeneity to functional roles.
In this regime, critical exponents no longer reflect topology alone, but encode how operational responsibilities are distributed across the system.
The resulting diversity of scaling behaviors signals a shift from purely structural criticality toward a functional notion of universality in higher-order networks.

More broadly, directed hypergraph percolation provides a unifying framework for studying robustness in systems governed by asymmetric, higher-order interactions.
By explicitly coupling topology, directionality, and functional constraints, it enables a principled characterization of resilience across biochemical, technological, and socio-economic domains.
Future extensions could incorporate temporal variability, adaptive rewiring, or dynamical processes unfolding on directed hypergraphs, allowing one to examine how functional criticality co-evolves with network structure.
Taken together, our results suggest that understanding robustness in complex systems requires moving beyond connectivity alone, toward the logic by which interactions are functionally realized.

\section*{MATERIALS AND METHODS}
\label{methods}

\subsection*{Construction of synthetic directed hypergraphs}

We generate ensembles of random directed hypergraphs using a configuration-model framework generalized to higher-order, directed interactions.
This approach enables independent control of node degree and hyperedge cardinality distributions while otherwise randomizing connections.

As a homogeneous baseline, we consider hypergraphs with fixed cardinalities $m^{\rm in}=3$ and $m^{\rm out}=2$.
Nodes are assigned to hyperedges uniformly at random, resulting in Poisson-distributed in- and out-degrees and minimal structural heterogeneity.

Uncorrelated scale-free hypergraphs are constructed by independently sampling power-law sequences $P(x)\sim x^{-\gamma}$ for $q^{\rm in}$, $q^{\rm out}$, $m^{\rm in}$, and $m^{\rm out}$ via inverse transform sampling.
Sequences are rescaled to satisfy stub-conservation constraints,
$\sum_i q_i^{\rm in}=\sum_\alpha m_\alpha^{\rm out}$ and $\sum_i q_i^{\rm out}=\sum_\alpha m_\alpha^{\rm in}$,
followed by random pairing of input and output stubs, yielding networks without degree–cardinality correlations.

Maximally correlated scale-free hypergraphs are generated by enforcing $q_i^{\rm in}=q_i^{\rm out}$ for nodes and $m_\alpha^{\rm in}=m_\alpha^{\rm out}$ for hyperedges, with both sequences drawn from a single power-law distribution.
This construction induces strong input–output correlations, producing networks in which hubs dominate both incoming and outgoing connectivity.

\subsection*{Construction of an empirical directed hypergraph}

We constructed a directed hypergraph representation of the \textit{Escherichia coli} metabolic network (model iJO1366).
Starting from the stoichiometric matrix $S$, metabolites were mapped to nodes and reactions to directed hyperedges.
For each reaction $j$, metabolites with negative stoichiometric coefficients ($S_{ij}<0$) were identified as inputs (substrates), whereas those with positive coefficients ($S_{ij}>0$) were identified as outputs (products), consistent with thermodynamic reaction directionality.
This procedure yields a connected directed hypergraph with $N=1805$ nodes and $M=2253$ hyperedges.

Anchor nodes were assigned deterministically based on biochemical constraints rather than stochastic rules.
For each hyperedge, all substrate metabolites were designated as anchors, reflecting the stoichiometric requirement that the depletion of any essential substrate suppresses the corresponding reaction.
This construction embeds functional dependencies directly into the hypergraph structure.

\subsection*{Finite-size scaling analysis}\label{sec11}

Finite-size scaling (FSS) was used to quantify the critical behavior of directed hypergraph percolation~\cite{radicchi2015breaking}.
The order parameter is the percolation strength $R^{(+)}$ or $R^{(-)}$, defined as the fraction of nodes in HGIN and HGOUT.
The corresponding susceptibility is
\begin{equation}
\chi(p_N,N)=N\bigl(\langle R^{2}\rangle-\langle R\rangle^{2}\bigr).
\end{equation}

For a finite system of size $N$, the order parameter follows the scaling form
\begin{equation}
R = N^{-\beta/\nu} F\!\left(|p_N-p_c|N^{1/\nu}\right),
\end{equation}
where $\beta$ and $\nu$ are the standard critical exponents.
The pseudo-critical threshold $p_c(N)$ is identified as the location of the susceptibility maximum and converges to the asymptotic critical point as
\begin{equation}
p_c(N)-p_c \propto N^{-1/\nu}.
\end{equation}

Evaluating the order parameter at $p=p_c(N)$ yields
$R(p_c(N))\propto N^{-\beta/\nu}$,
allowing direct estimation of exponent ratios.

Since theory predicts that the HGIN, HGOUT, and HGSCC emerge at the same critical point, a unified FSS protocol was adopted.
For each system size $N$, a single pseudo-critical point $p_c(N)$ was determined from the susceptibility peak of the HGIN or HGOUT, and all three order parameters $R^{(+)}$, $R^{(-)}$, and $R$ were evaluated at this common value.

The exponent $\nu$ was obtained from the finite-size drift of $p_c(N)$, while $\beta^{(+)}/\nu$, $\beta^{(-)}/\nu$, and $\beta/\nu$ were extracted from the size dependence of the corresponding order parameters (Fig.~\ref{fig:FSS}).
Combining these estimates yields $\beta^{(+)}\approx1.1$, $\beta^{(-)}\approx1.0$, and $\beta\approx2.0$, consistent with the analytical predictions.

%%%%%%%%%%%%%%%% REFERENCES %%%%%%%%%%%%%%%

\clearpage % Clear all remaining figures and tables then start a new page

% The list of references goes after the main text and before the acknowledgements
% When preparing an initial submission, we recommend you use BibTeX, like this:
%
\bibliography{bibliography} % for a file named science_template.bib
\bibliographystyle{sciencemag}

% After the paper has completed peer review and been revised ready for acceptance,
% you should comment out the lines above and copy-paste the contents of your .bbl
% file here instead. This will help ensure that our conversion software works correctly.
% Remember to re-run BibTeX first - check the timestamp!
%
% Example of the first three entries copy-pasted from science_template.bbl:
%
%\begin{thebibliography}{1}
%
%\bibitem{example}
%A.~N. {Author}, An example reference. \emph{Journal of Improbable Research}
%  \textbf{1}, 67 (2020).
%
%\bibitem{example2}
%F.~M. {Surname}, S.~{Author}, A second example. \emph{Interesting Research
%  Letters} \textbf{32}, 897 (2019).
%
%\bibitem{example_preprint}
%P.~{One}, P.~{Two}, P.~{Three}, {An unpublished preprint}. \emph{preprint}
%  (2021), arXiv:2101.12345.
%
%\end{thebibliography}

%%%%%%%%%%%%%%%% ACKNOWLEDGEMENTS %%%%%%%%%%%%%%%

\section*{Acknowledgments}
%Here you can thank helpful colleagues who did not meet the journal's authorship criteria, or provide other acknowledgments that don't fit the (compulsory) subheadings below. Formatting requirements for each of these sections differ between the \textit{Science}-family journals; consult the instructions to authors on the journal website for full details.
\paragraph*{Funding:}
This research was supported by the National Natural Science Foundation of China (T2422010 and 62172170) and “the Fundamental Research Funds for Central Universities”.
\paragraph*{Author contributions:}
X.L. and G.B. conceived the study and designed the overall research plan.  
Y.S., X.L. and G.B. developed the theoretical framework and carried out the analytical derivations.  
Y.S. performed the numerical simulations and data analysis.  
X.L. and G.B. supervised the project and contributed to the interpretation of the results.  
All authors discussed the findings, contributed to the manuscript preparation, and approved the final version.
\paragraph*{Competing interests:}
The authors declare that they have no competing interests.
\paragraph*{Data and materials availability:}
All data needed to evaluate the conclusions in the paper are present in the paper and/or the Supplementary Materials. The processed network files and the source code used for the percolation on directed hypergraph have been deposited in a public repository on GitHub at \url{https://github.com/yunxuesun/directed_hypergraph_percolation}.

% The DNC email network data can be accessed from the Network Repository \url{https://networkrepository.com/}.

\newpage

% --- 封面页 ---
\begin{center}
    {\Large Supplementary Materials for} \\
    \vspace{1em}
    {\fontsize{14pt}{16pt}\selectfont \textbf{Directionality and node heterogeneity reshape criticality in \\hypergraph percolation}} \\
    \vspace{1em}
    
    % 作者：3人以上只写一作 + et al. (12pt Roman)
    {\fontsize{12pt}{14pt}\selectfont Yunxue Sun \textit{et al.}} \\
    \vspace{1em}
    
    % 通讯作者
    {\fontsize{12pt}{14pt}\selectfont Corresponding author: Xueming Liu, xm\_liu@hust.edu.cn; \\Ginestra Bianconi, ginestra.bianconi@gmail.com} \\
\end{center}

\vspace{2em}

% Author list for the supplement
% Indicate the corresponding authors, but do NOT include institutions here
% It would be nice if the template auto-generated this, but doing so is complicated...
% Yunxue Sun,
% Xueming Liu$^{\ast}$,
% Ginestra Bianconi$^{\ast}$\\ % we're not in a \author{} environment this time, so use \\ for a new line
% \small$^\ast$Corresponding author. Email: xm\_liu@hust.edu.cn; ginestra.bianconi@gmail.com\\

\vspace{1em}
\noindent \textbf{This PDF file includes:}

\begin{itemize}
    \item[] Supplementary Text
    \item[] Figs. S1 to S6
\end{itemize}

\newpage
\setcounter{equation}{0}
\renewcommand\theequation{{S\arabic{equation}}}
\renewcommand\thetable{{S\Roman{table}}}
\renewcommand\thefigure{{S\arabic{figure}}}
\setcounter{equation}{0}
\setcounter{figure}{0}
\setcounter{section}{0}
\renewcommand{\thesection}{S\arabic{section}}
\section{Derivation of the message passing algorithm}
\label{ApA}
In this section, we derive the message-passing theory presented in the main text. We begin by formulating the exact message-passing equations for a specific realization of damage on a fixed hypergraph, where full information about the initial state of all nodes and hyperedges is known.

Let us assume that we know whether each node $i$ is initially damaged ($x_i=0$) or intact ($x_i=1$), and whether each hyperedge $\alpha$ is initially damaged ($y_{\alpha}=0$) or intact ($y_{\alpha}=1$). Moreover, as mentioned in the main text, for each hyperedge $\alpha$, we denote by $z_{i\alpha}$ the condition where node $i$ serves as an anchor node, setting $z_{i\alpha}=0$ otherwise.
Note that the variable $y_{\alpha}$ does not take into account the damage induced by the removal of anchor nodes. 

The message passing algorithm implementing directed hypergraph percolation with anchor nodes consists in updating the forward and backward messages $\hat{w}_{i\rightarrow \alpha}^{(\pm)}$  and $\hat{v}_{\alpha\to i}^{(\pm)}$ going from node $i$ to hyperedge $\alpha$ and from hyperedge $\alpha$ to node $i$.
These message passing equations extend previous approaches adopted in undirected hypergraphs to directed hypergraphs \cite{Bianconi2024Nature,2025percolation}.
The message $\hat{w}_{i\to\alpha}^{(+)}$ is equal to one if  all the following conditions are satisfied:

\begin{enumerate}[label=(\roman*), leftmargin=0pt, itemindent=3em]
%\item 
%node $i$ is not initially damaged, i.e. $x_i=1$, 
\item 
node $i$ is either an anchor node (i.e., $z_{i\alpha}=1$) or, if it is not an anchor node (i.e., $z_{i\alpha}=0$), node $i$ is not initially damaged.
\item 
node $i$ receives at least one positive forward message $\hat{v}_{\beta\to i}^{(+)}=1$ from at least one of the hyperedges $\beta$ that point to it.
\end{enumerate}

In all the other cases $\hat{w}_{i\to\alpha}^{(+)}=0$. 

Similarly, the message $\hat{v}_{\alpha\to i}^{(+)}$ is equal to one if all the following conditions are satisfied:

\begin{enumerate}[label=(\roman*), leftmargin=0pt, itemindent=3em]
\item 
the hyperedge $\alpha$ is not initially damaged (i.e., $y_{\alpha}=1$);

\item 
all anchor nodes $j$ belonging to hyperedge $\alpha$ (where $z_{j\alpha}=1$), distinct from node $i$, are intact;

\item 
the hyperedge $\alpha$ receives at least one positive forward message $\hat{w}_{j\to \alpha}^{(+)}=1$ from at least one incoming node $j$ distinct from $i$.
\end{enumerate}

In all the other cases $\hat{v}_{\alpha\to i}^{(+)}=0.$

By defining the algorithm determining the backward messages $\hat{w}_{i\to \alpha}^{(-)}$ and $\hat{v}_{\alpha\to i}^{(-)}$ in an analogous manner, we obtain the complete message-passing framework for directed hypergraph percolation with anchor nodes
\begin{align}
\hat{w}_{i\rightarrow \alpha}^{(+)} &= x_i^{1-z_{i\alpha}} \left[ 1 - \prod_{\beta\in N^{(-)}(i)\setminus \alpha}(1 - \hat{v}_{\beta\rightarrow i}^{(+)}) \right],\nonumber\\
\hat{v}_{\alpha\to i}^{(+)} &=
y_{\alpha}\left(\prod_{j\in N(\alpha)\setminus i}\!x_j^{z_{j\alpha}}\!\right)\!\!\ \left[ 1 - \!\prod_{j\in N^{(-)}(\alpha)\setminus i}\!(1 - \hat{w}_{j\rightarrow \alpha}^{+}) \right],\nonumber\\
\hat{w}_{i\rightarrow \alpha}^{(-)} &=  x_i^{1-z_{i\alpha}}\left[ 1 - \prod_{\beta\in N^{(+)}(i)\setminus \alpha}(1 - \hat{v}_{\beta\rightarrow i}^{(-)}) \right]
,\nonumber\\
\hat{v}_{\alpha\to i}^{(-)} &=
y_{\alpha}\left(\prod_{j\in N^(\alpha)\setminus i}\!x_j^{z_{j\alpha}}\!\right)\!\!\ \left[ 1 - \!\prod_{j\in N^{(+)}(\alpha)\setminus i}\!(1 - \hat{w}_{j\rightarrow \alpha}^{(-)}) \right].
\label{mp_1aa}
\end{align}
where $N(\alpha)$ indicates the set of node belonging to hyperedge $\alpha$.

The indicator function $\hat{r}_i^{(+)}$ takes the value one, indicating that node $i$ belongs to the HGOUT component, if and only if:

\begin{enumerate}[label=(\roman*), leftmargin=0pt, itemindent=3em]
\item 
node $i$ is not initially damaged (i.e., $x_i=1$); 

\item 
node $i$ receives at least one positive forward message $\hat{v}_{\alpha\to i}^{(+)}=1$ from at least one of the incoming hyperedges $\alpha$.
\end{enumerate}

Similarly, the indicator function $\hat{s}_\alpha^{(+)}$ takes the value one, indicating that hyperedge $\alpha$ belongs to the HGOUT component, if and only if:

\begin{enumerate}[label=(\roman*), leftmargin=0pt, itemindent=3em]
\item 
the hyperedge $\alpha$ is not initially damaged (i.e., $y_\alpha=1$); 

\item 
all anchor nodes $j$ belonging to hyperedge $\alpha$ are intact;

\item 
the hyperedge $\alpha$ receives at least one positive forward message $\hat{w}_{i\to \alpha}^{(+)}=1$ from at least one incoming node $i$. 
\end{enumerate}

This algorithm, along with the corresponding one for $\hat{r}_i^{(-)}$ and $\hat{s}_{\alpha}^{(-)}$, is encoded in the following equations
\begin{align}
\hat{r}_{i}^{(+)} &=  x_i^{1-z_{i\alpha}}\left[ 1 - \prod_{\alpha\in N^{(-)}(i)}(1 - v_{\alpha\rightarrow i}^{(+)}) \right], \nonumber\\
\hat{s}_{\alpha}^{(+)} &= y_{\alpha}\left(\prod_{j\in N(\alpha)\setminus i}x_j^{z_{j\alpha}}\right)\left[ 1 - \prod_{i\in N^{(-)}(\alpha)}(1 - w_{i\rightarrow \alpha}^{(+)}) \right],  \nonumber\\
\hat{r}_{i}^{(-)} &=  x_i^{1-z_{i\alpha}}\left[ 1 - \prod_{\alpha\in N^{(+)}(i)}(1 - v_{\alpha\rightarrow i}^{(-)}) \right], \nonumber\\
\hat{s}_{\alpha}^{(-)} &= y_{\alpha}\left(\prod_{j\in N(\alpha)\setminus i}x_j^{z_{j\alpha}}\right)\left[ 1 - \prod_{i\in N^{(+)}(\alpha)}(1 - w_{i\rightarrow \alpha}^{(-)}) \right].
\end{align}

Finally, the indicator functions $\hat{r}_i$ and $\hat{s}_{\alpha}$ determining whether node $i$ and hyperedge $\alpha$ belong to the HGSCC are calculated by requiring that they belong to both the HGIN and HGOUT components.

Specifically, the indicator function $\hat{r}_i$ takes the value one if and only if:

\begin{enumerate}[label=(\roman*), leftmargin=0pt, itemindent=3em]
\item 
node $i$ is not initially damaged (i.e., $x_i=1$); 

\item 
node $i$ receives at least one positive forward message $\hat{v}_{\alpha\to i}^{(+)}=1$ from at least one of the hyperedges $\alpha$ that point to it;

\item 
node $i$ receives at least one positive backward message $\hat{v}_{\alpha\to i}^{(-)}=1$ from at least one of the hyperedges $\alpha$ pointed to by node $i$.
\end{enumerate}

The indicator function $\hat{s}_\alpha$ determining whether hyperedge $\alpha$ belongs to the HGSCC takes the value one if and only if:

\begin{enumerate}[label=(\roman*), leftmargin=0pt, itemindent=3em]
\item 
the hyperedge $\alpha$ is not initially damaged; 

\item 
all anchor nodes $j$ belonging to hyperedge $\alpha$ are intact;

\item 
the hyperedge $\alpha$ receives at least one positive forward message $\hat{w}_{i\to \alpha}^{(+)}=1$ from at least one incoming node $i$;

\item 
the hyperedge $\alpha$ receives at least one positive backward message $\hat{w}_{j\to \alpha}^{(-)}=1$ from at least one outgoing node $j$ (i.e., a node pointed to by hyperedge $\alpha$).
\end{enumerate}

Thus, the indicator functions $\hat{r}_i$ and $\hat{s}_{\alpha}$ for the HGSCC are given by

\begin{eqnarray} 
\hat{r}_{i} &=&  x_i\left[ 1 - \prod_{\alpha\in N^{(-)}(i)}(1 - v_{\alpha\rightarrow i}^{(+)}) \right]\left[ 1 - \prod_{\alpha\in N^{(+)}(i)}(1 - v_{\alpha\rightarrow i}^{(-)}) \right],\nonumber \\
\hat{s}_{\alpha} &=& y_{\alpha}\left(\prod_{j\in N(\alpha)\setminus i}x_j^{z_{j\alpha}}\right)\left[ 1 - \prod_{i\in N^{(-)}(\alpha)}(1 - w_{i\rightarrow \alpha}^{(+)}) \right]\left[ 1 - \prod_{i\in N^{(+)}(\alpha)}(1 - w_{i\rightarrow \alpha}^{(-)}) \right].
\label{hmp_1b}
\end{eqnarray}

These are the fundamental message-passing equations valid when the full configuration of node and hyperedge damage, $\{x_i\}=\{x_1,x_2,\ldots, x_N\}$ and $\{y_{\alpha}\}=\{y_1,y_2,\ldots, y_M\}$, is exactly known.

However, in many scenarios, we have access only to partial information, assuming that nodes and hyperedges of cardinality ${\bf m}$ are damaged independently at random with probabilities $p_N$ and $p_H^{[{\bf m}]}$, respectively. In this case, although the exact configuration of the initial damage is unknown, its associated probability distribution is given by
\begin{align}
\mathbb{P}(\{x_i\},\{y_{\alpha}\}) &= \left[\prod_{i=1}^N p_N^{x_i}(1-p_N)^{1-x_i}\right] 
\left[\prod_{\alpha=1}^M \Big(p_H^{[{\bf m}_{\alpha}]}\Big)^{y_\alpha}\Big(1-p_H^{[{\bf m}_{\alpha}]}\Big)^{1-y_\alpha}\right].
\label{Pxy}
\end{align}

Under these conditions, we can derive the message-passing theory presented in main text, expressed in terms of the messages $w_{i\to\alpha}^{(\pm)}$ and $v^{(\pm)}_{\alpha\to i}$, by averaging the stochastic messages $\hat{w}_{i\to\alpha}^{(\pm)}$ and $\hat{v}^{(\pm)}_{\alpha\to i}$ over the distribution $\mathbb{P}(\{x_i\},\{y_{\alpha}\})$. This derivation follows the general approach outlined in Ref. \cite{2018multilayer}, which has previously been applied to percolation on undirected hypergraphs in \cite{Bianconi2024Nature,2025percolation}.

Similarly, the probabilities $r_i^{(\pm)}$ and $s_{\alpha}^{(\pm)}$ that nodes/hyperedges belong to the HGOUT/HGIN components are obtained by averaging the indicator functions $\hat{r}_i^{(\pm)}$ and $\hat{s}_{\alpha}^{(\pm)}$. Finally, the probabilities $r_i$ and $s_{\alpha}$ that nodes/hyperedges belong to the HGSCC are obtained by averaging the indicator functions $\hat{r}_i$ and $\hat{s}_{\alpha}$
\begin{equation}
\begin{split}
r_{i}^{(+)} &= p_N \left[ 1 - \prod_{\alpha\in N^{(-)}(i)}(1 - v_{\alpha \rightarrow i}^{(+)}) \right], \\
s_{\alpha}^{(+)} &= p_H^{[{\bf m}_{\alpha}]}p_N^{Z_{\alpha}}\left[ 1 - \prod_{i\in N^{(-)}(\alpha)}(1 - w_{i\rightarrow \alpha}^{(+)}) \right], \\
r_{i}^{(-)} &= p_N \left[ 1 - \prod_{\alpha\in N^{(+)}(i)}(1 - v_{\alpha \rightarrow i}^{(-)}) \right], \\
s_{\alpha}^{(-)} &= p_H^{[{\bf m}_{\alpha}]}p_N^{Z_{\alpha}}\left[ 1 - \prod_{i\in N^{(+)}(\alpha)}(1 - w_{i\rightarrow \alpha}^{(-)}) \right],
\end{split}
\label{sp_1b}
\end{equation}
where $Z_{\alpha}=\sum_{i\in N(\alpha)} z_{i\alpha}$.
\section{Simultaneous Emergence of Giant Components}
\label{sec:S3}

The critical behavior of percolation in directed hypergraphs is controlled by the linear stability of the underlying message-passing equations. The appearance of macroscopic connectivity corresponds to a continuous phase transition where the trivial fixed point loses stability. 

First, the onset of the HGOUT is governed by the linearization of the forward propagation dynamics around this vacuum state
\begin{equation}
\begin{split}
w_{i\to\alpha}^{(+)}&=p_N^{1 - z_{i\alpha}}\sum_{\beta\in N^{(-)}(i)\setminus \alpha} v_{\beta\to i}^{(+)},
\\
v_{\alpha\to i}^{(+)}&=p_H^{[{\bf m}_{\alpha}]}p_N^{Z_{i\alpha}}\sum_{j\in N^{(-)}(\alpha)\setminus i} w_{j\to \alpha}^{(+)}.
\end{split}
\end{equation}
This system of linear equations admits a non-zero solution if and only if the leading eigenvalue, $\Lambda^{(+)}(p_N,{\bf p}_H)$, of the associated Jacobian matrix exceeds unity. This Jacobian corresponds to the hypergraph non-backtracking matrix $\mathcal{A}^{(+)}$, which possesses a block structure defined on the directed bipartite graph
\begin{equation}
\mathcal{A}^{(+)}=\left(\begin{array}{ll}0 &\mathcal{B}^{NH (+)}\\
\mathcal{B}^{HN (+)}& 0\end{array}\right).
\end{equation}
Here, the blocks $\mathcal{B}^{NH(+)}$ and $\mathcal{B}^{HN(+)}$ describe the propagation of influence from nodes to hyperedges and vice versa. Their elements are explicitly given by
\begin{equation}
\begin{split}
\mathcal{B}^{NH (+)}_{(i\to \alpha);(\beta\to j)}&=p_N^{1 - z_{i\alpha}}\delta_{j,i}(1-\delta_{\beta,\alpha}),
\\
\mathcal{B}^{HN (+)}_{(\alpha\to i);( j\to \beta)}&=p_H^{[{\bf m}_{\alpha}]}p_N^{Z_{i\alpha}}\delta_{\alpha,\beta}(1-\delta_{i,j}),
\label{3800}
\end{split}
\end{equation}
where $\delta_{x,y}$ indicates the Kronecker delta. Consequently, the critical point for the emergence of the HGOUT is determined by the condition $\Lambda^{(+)}(p_N,{\bf p}_H)=1$.

By symmetry, an analogous stability criterion governs the emergence of the HGIN. Although the backward messages propagate against the direction of the hyperedges, the critical point is similarly determined by the linear stability of the trivial fixed point of the backward update rules. This yields the condition
\begin{equation}
    \Lambda^{(-)}(p_N, p_H) > 1,
    \label{eq:S-10}
\end{equation}
where $\Lambda^{(-)}$ is the leading eigenvalue of the backward non-backtracking matrix $\mathcal{A}^{(-)}$. This matrix possesses a block structure isomorphic to the forward case but defined by the reversed flow of influence
\begin{equation}
    \mathcal{A}^{(-)} = \begin{pmatrix} 0 & \mathcal{B}^{NH(-)} \\ \mathcal{B}^{HN(-)} & 0 \end{pmatrix},
    \label{eq:S-11}
\end{equation}
with the elements explicitly given by
\begin{equation} 
\begin{split}
    \mathcal{B}_{(i\to\alpha);(\beta\to j)}^{NH(-)} &= p_N^{1-z_{i\alpha}} \delta_{j,i} (1-\delta_{\beta,\alpha}), \\
    \mathcal{B}_{(\alpha\to i);(j\to\beta)}^{HN(-)} &= p_H^{[{\bf m}_\alpha]} p_N^{Z_{i\alpha}} \delta_{\alpha,\beta} (1-\delta_{i,j}).
\end{split}
    \label{eq:S-12}
\end{equation}

Crucially, despite the directional inversion, the structural symmetry of the random hypergraph ensemble implies that the spectral properties of $\mathcal{A}^{(-)}$ and $\mathcal{A}^{(+)}$ are identical. Consequently, the leading eigenvalues satisfy $\Lambda^{(+)} = \Lambda^{(-)}$, strictly ensuring that the HGIN and HGOUT components emerge simultaneously at the same percolation threshold, which also marks the onset of the HGSCC.

\section{Mean-Field Theory and Percolation Thresholds}
\label{si-aver}
Averaging the message-passing algorithm over the anchor node distribution requires careful treatment. Specifically, one must ensure that the average over the random variable $z_{i\alpha}$ is not performed independently on two subsequent messages that are correlated. To address this, we introduce the auxiliary messages $\omega_{i\rightarrow\alpha}^{(\pm)}$ by expressing the dependence on the anchor state explicitly. We define the relationship
\bea
w_{i\rightarrow\alpha}^{(\pm)}={p}_N^{1-z_{i\alpha}}\omega_{i\rightarrow\alpha}^{(\pm)},
\eea

Substituting this expression into Eq. \ref{m_anchor_e} yields the following update rules for the auxiliary messages

\begin{equation}
\label{m_anchor2_e} 
\begin{aligned} 
\omega_{i\rightarrow \alpha}^{(+)} &= \left[ 1 - \prod_{\beta\in N^{(-)}(i)\setminus \alpha}(1 - v_{\beta\rightarrow i}^{(+)}) \right], \\
v_{\alpha\rightarrow i}^{(+)} &= p_H^{[{\bf m}_{\alpha}]} p_N^{Z_{i\alpha}}\left[ 1 - \prod_{j\in N^{(-)}(\alpha)\setminus i}(1 - {p}_N^{1-z_{i\alpha}}\omega_{j\rightarrow \alpha}^{(+)}) \right], \\
\omega_{i\rightarrow \alpha}^{(-)} &=   \left[ 1 - \prod_{\beta\in N^{(+)}(i)\setminus \alpha}(1 - v_{\beta\rightarrow i}^{(-)}) \right], \\
v_{\alpha\rightarrow i}^{(-)} &= p_H^{[{\bf m}_{\alpha}]} p_N^{Z_{i\alpha}}\left[ 1 - \prod_{j\in N^{(+)}(\alpha)\setminus i}(1 - {p}_N^{1-z_{i\alpha}}\omega_{j\rightarrow \alpha}^{(-)}) \right].
\end{aligned}
\end{equation}

By performing the average of these messages over the anchor node distribution, given by
\bea
\mathbb{P}(\{z_{i\alpha}\})=\prod_{i\alpha}\theta^{z_{i\alpha}}(1-\theta)^{1-z_{i\alpha}}.
\eea
we derive the self-consistent equations for the average messages $\tilde{\omega}_{i\rightarrow \alpha}^{(\pm)}$ and $\tilde{v}_{\alpha\rightarrow i}^{(\pm)}$
\begin{equation}
\label{m_anchor3_e} 
\begin{aligned} 
\tilde{\omega}_{i\rightarrow \alpha}^{(+)} &= \left[ 1 - \prod_{\beta\in N^{(-)}(i)\setminus \alpha}(1 - \tilde{v}_{\beta\rightarrow i}^{(+)}) \right], \\
\tilde{v}_{\alpha\rightarrow i}^{(+)} &= p_H^{[{\bf m}_{\alpha}]} \pi_N^{m_{\alpha}-1}\left[ 1 - \prod_{j\in N^{(-)}(\alpha)\setminus i}(1 - \hat{p}_N\tilde{\omega}_{j\rightarrow \alpha}^{(+)}) \right], \\
\tilde{\omega}_{i\rightarrow \alpha}^{(-)} &=   \left[ 1 - \prod_{\beta\in N^{(+)}(i)\setminus \alpha}(1 - \tilde{v}_{\beta\rightarrow i}^{(-)}) \right], \\
\tilde{v}_{\alpha\rightarrow i}^{(-)} &= p_H^{[{\bf m}_{\alpha}]} \pi_N^{m_{\alpha}-1}\left[ 1 - \prod_{j\in N^{(+)}(\alpha)\setminus i}(1 - \hat{p}_N\omega_{j\rightarrow \alpha}^{(-)}) \right].
\end{aligned}
\end{equation}
where $\pi_N$ and $\hat{p}_N$ are given by 
\begin{equation}
\begin{aligned} 
\pi_N&=1-\theta+\theta p_N,\\
\hat{p}_N&=p_N/\pi_N.
\end{aligned}
\end{equation}
In terms of these messages, the probabilities $r_i^{(\pm)}$ and $s_{\alpha}^{(+)}$ are given by 
\begin{equation}
\begin{aligned}
r_{i}^{(+)} &= p_N \left[ 1 - \prod_{\alpha\in N^{(-)}(i)}(1 - \tilde{v}_{\alpha \rightarrow i}^{(+)}) \right], \nonumber\\
s_{\alpha}^{(+)} &= p_H^{[{\bf m}_{\alpha}]}\pi_N^{m_{\alpha}}\left[ 1 - \prod_{i\in N^{(-)}(\alpha)}(1 - \hat{p}_N\tilde{\omega}_{i\rightarrow \alpha}^{(+)}) \right], \nonumber\\
r_{i}^{(-)} &= p_N \left[ 1 - \prod_{\alpha\in N^{(+)}(i)}(1 - \tilde{v}_{\alpha \rightarrow i}^{(-)}) \right], \nonumber\\
s_{\alpha}^{(-)} &= p_H^{[{\bf m}_{\alpha}]}\pi_N^{m_{\alpha}}\left[ 1 - \prod_{i\in N^{(+)}(\alpha)}(1 - \hat{p}_N\tilde{\omega}_{i\rightarrow \alpha}^{(-)}) \right].
\label{sp_1b3}
\end{aligned}
\end{equation}

Similarly, we obtain the expressions for $r_{i}$ and $s_{\alpha}$ as
\begin{equation}
\begin{aligned}
r_{i} = &p_N \left[ 1 - \prod_{\alpha\in N^{(+)}(i)}(1 - \tilde{v}_{\alpha \rightarrow i}^{(-)}) \right]\left[ 1 - \prod_{\alpha\in N^{(-)}(i)}(1 - \tilde{v}_{\alpha \rightarrow i}^{(+)}) \right],\\
s_{\alpha} =& p_H^{[{\bf m}_{\alpha}]}\pi_N^{m_{\alpha}}\left[ 1 - \prod_{i\in N^{(-)}(\alpha)}(1 - \hat{p}_N\tilde{\omega}_{i\rightarrow \alpha}^{(+)}) \right] \left[ 1 - \prod_{i\in N^{(+)}(\alpha)}(1 - \hat{p}_N\tilde{\omega}_{i\rightarrow \alpha}^{(-)}) \right].
\end{aligned}
\end{equation}

To extract generic behavior beyond specific realizations, we extend the analysis to an ensemble of random directed hypergraphs specified by a joint node degree distribution $P(\mathbf{q})$ and a hyperedge cardinality distribution $Q(\mathbf{m})$. Within this mean-field framework, anchor-node heterogeneity is incorporated by assuming that, for each hyperedge, its constituent nodes are independently assigned as anchors with probability $\theta$.

To describe percolation on this random ensemble~\cite{courtney2016generalized,Chodrow2021Hypergraph}, we introduce two sets of order parameters: $W^{(\pm)}$ denotes the probability that a randomly chosen node connects to the giant components through one of its incident hyperedges, while $V^{(\pm)}$ denotes the probability that a hyperedge propagates connectivity to an attached node. Averaging the microscopic message-passing equations (Eq. \ref{m_anchor3_e}) over the network ensemble yields the following system of self-consistent equations:

\begin{equation}
\begin{split}
\hspace{-10mm}W^{(+)} &=  \sum_{\bf q} \frac{q^{\rm out}P({\bf q})}{\avg{q^{\rm out}}} \left[ 1 - (1 - V^{(+)})^{q^{\rm in}} \right],
\\[3pt]
\hspace{-10mm}V^{(+)} &= \sum_{{\bf m}} p_H^{[{\bf m}]} {\pi}_N^{m-1}\frac{m^{\rm out}Q({\bf m})}{\avg{m^{\rm out}}} \left[ 1 - (1 - \hat{p}_NW^{(+)})^{m^{\rm in}
} \right],\\
\hspace{-10mm}W^{(-)} &=  \sum_{\bf q} \frac{q^{\rm in}P({\bf q})}{\avg{q^{\rm in}}} \left[ 1 - (1 - V^{(-)})^{q^{\rm out}} \right],\\[3pt]
\hspace{-10mm}V^{(-)} &= \sum_{{\bf m}} p_H^{[{\bf m}]} {\pi}_N^{m-1}\frac{m^{\rm in}Q({\bf m})}{\avg{m^{\rm in}}} \left[ 1 - (1 - \hat{p}_NW^{(-)})^{m^{\rm out}} \right],
\label{self}
\end{split}
\end{equation}
where $\pi_N = 1 - \theta + \theta p_N$ represents the average probability that a node is either a non-anchor or an intact anchor. Consequently, $\hat{p}_N = p_N/\pi_N$ corresponds to the effective survival probability renormalized by the anchor constraint.

Once the message-passing process stabilizes, the macroscopic sizes of the giant components are obtained by averaging the connectivity probabilities over the full ensemble. The relative sizes of the directional components are given by
\begin{equation}
\begin{split} 
R^{(+)} &= p_N \sum_{\bf q} P({\bf q}) \left[ 1 - (1 - V^{(+)})^{q^{\rm in}} \right]
,\\
S^{(+)} &= \sum_{\bf m}
p_H^{[{\bf m}]}\pi_N^{m}Q({\bf m}) \left[ 1 - (1 - \hat{p}_NW^{(+)})^{m^{\rm in}} \right],\\
R^{(-)} &= p_N \sum_{\bf q} P({\bf q}) \left[ 1 - (1 - V^{(-)})^{q^{\rm out}} \right],\\
S^{(-)} &= \sum_{\bf m} p_H^{[{\bf m}]}\pi_N^{m}Q({\bf m}) \left[ 1 - (1 -\hat{p}_N W^{(-)})^{m^{\rm out}} \right].
\label{90b}
\end{split}
\end{equation}
Correspondingly, since the HGSCC represents the intersection of the HGIN and HGOUT, the fractions of nodes $R$ and hyperedges $S$ belonging to it are determined by
\begin{equation}
\begin{split} 
R &= p_N \sum_{\bf q} P({\bf q}) \left[ 1 - (1 - V^{(-)})^{q^{\rm out}} \right]\left[ 1 - (1 - V^{(+)})^{q^{\rm in}} \right], \\
S &= \sum_{\bf m} p_H^{[{\bf m}]}\pi_N^{m}Q({\bf m}) \left[ 1 - (1 -\hat{p}_N W^{(+)})^{m^{\rm out}} \right]\left[ 1 - (1 - \hat{p}_NW^{(-)})^{m^{\rm in}} \right].
\label{SRh}
\end{split}
\end{equation}

The percolation threshold is determined by the linear stability of the trivial fixed point ($W^{(\pm)}=0, V^{(\pm)}=0$) of the self-consistent equations. To locate this transition, we expand Eq.~\ref{self} to first order around zero
\begin{equation}
\begin{split}
W^{(+)} &\approx \frac{\langle q^{\rm in}q^{\rm out}\rangle}{\langle q^{\rm out}\rangle} V^{(+)}, \\
V^{(+)} &\approx \frac{\langle p_H^{[\mathbf{m}]}\pi_N^{m-2}m^{\rm in}m^{\rm out}\rangle}{\langle m^{\rm out}\rangle} \hat{p}_N W^{(+)}.
\end{split}
\end{equation}
Substituting the expression for $V^{(+)}$ into that for $W^{(+)}$ yields a closed-form condition for the onset of instability
\begin{equation}
W^{(+)} \approx \left[ \hat{p}_N \frac{\langle q^{\rm in}q^{\rm out}\rangle}{\langle q^{\rm out}\rangle} \frac{\langle p_H^{[\mathbf{m}]}\pi_N^{m-2}m^{\rm in}m^{\rm out}\rangle}{\langle m^{\rm out}\rangle} \right] W^{(+)}.
\end{equation}

The critical point corresponds to the divergence of the susceptibility, occurring when the bracketed coefficient reaches unity. By recalling that $\hat{p}_N = p_N/\pi_N$ and absorbing the factor $\pi_N^{-1}$ into the second moment term, we arrive at the explicit criterion for the simultaneous emergence of the HGIN, HGOUT, and HGSCC
\begin{equation}
\hat{\Lambda} = {p}_N \frac{\langle q^{\rm in}q^{\rm out}\rangle}{\langle q^{\rm out}\rangle} \frac{\langle {p}_H^{[\bf m]}{\pi}_N^{m-2}m^{\rm out}m^{\rm in}\rangle}{\langle m^{\rm out}\rangle} = 1.
\label{thre_sm}
\end{equation}

Crucially, Eq.~\ref{thre_sm} reveals that anchor nodes impose a strictly stronger condition for global connectivity compared to the anchor-free case. The term $\pi_N^{m-2}$ systematically suppresses the contribution of large hyperedges, thereby elevating the percolation threshold. This behavior mirrors findings in undirected hypergraphs~\cite{Bianconi2024Theory,2025percolation}, confirming that the fragility driven by node heterogeneity is a universal feature extending to the directed case. Furthermore, for uncorrelated degree and cardinality distributions with finite first moments, the branching factor $\hat{\Lambda}$ remains finite, guaranteeing that the giant components emerge at a non-trivial critical point.

To highlight the symmetry breaking induced by anchor nodes, we consider the anchor-free limit. In this case, $\pi_N = 1$, and the general criterion simplifies to the standard condition
\begin{equation}
 p_N \frac{\langle q^{\rm in}q^{\rm out}\rangle}{\langle q^{\rm out}\rangle} \frac{\langle p_H^{[\mathbf{m}]}m^{\rm in}m^{\rm out}\rangle}{\langle m^{\rm out}\rangle} = 1.
\label{eq:anchor_free_general}
\end{equation}
Assuming uniform hyperedge reliability $p_H^{[\mathbf{m}]} = p_H$, this becomes
\begin{equation}
p_N p_H  \frac{\langle q^{\rm in}q^{\rm out}\rangle}{\langle q^{\rm out}\rangle} \frac{\langle m^{\rm in}m^{\rm out}\rangle}{\langle m^{\rm out}\rangle} = 1.
\label{eq:anchor_free_uniform}
\end{equation} 

In this limit, the system exhibits a fundamental symmetry: node and hyperedge survival probabilities play interchangeable roles. Consequently, the node percolation threshold and hyperedge percolation threshold coincide. However, the presence of anchor nodes breaks this symmetry, decoupling the two transitions. 
For pure hyperedge percolation, we have $\pi_N = 1$, and the threshold remains identical to the anchor-free scenario
\begin{equation}
 p_H  \frac{\langle q^{\rm in}q^{\rm out}\rangle}{\langle q^{\rm out}\rangle} \frac{\langle m^{\rm in}m^{\rm out}\rangle}{\langle m^{\rm out}\rangle} = 1.
\label{eq:hyperedge_perc}
\end{equation}
In stark contrast, for node percolation, the threshold is explicitly shifted by the anchor constraint. Substituting $\pi_N$ into the criterion shows that the critical point satisfies
\begin{equation}
 {p}_N \frac{\langle q^{\rm in}q^{\rm out}\rangle}{\langle q^{\rm out}\rangle} \frac{\langle (1-\theta+\theta p_N)^{m-2}m^{\rm in}m^{\rm out}\rangle}{\langle m^{\rm out}\rangle} = 1,
 \label{eq:node_perc}
\end{equation}
demonstrating that the node percolation threshold is strictly higher than that of hyperedges.

\section{Derivation of the critical indices}
\label{Apc}
In this section, we detail the derivation of the critical exponents for directed hypergraph percolation. We perform an asymptotic analysis of the self-consistent equations derived in the main text, expanding the message-passing probabilities near the critical point where the order parameters are small ($W^{(\pm)} \ll 1, V^{(\pm)} \ll 1$). We consider three distinct cases: (i) well-behaved distributions with finite moments, (ii) uncorrelated power-law distributions, and (iii) maximally correlated power-law distributions. For a review on asymptotic methods to calculate anomalous critical exponents for dynamics on finite scale and scale-free networks, we refer the reader to classical texts in network theory such as \cite{mendes2003evolution,Dorogovtsev2008Critical}.

\subsection{Case of well-behaved degree and cardinality distributions}

We first analyze the case where the degree distribution $P(\mathbf{q})$ and cardinality distribution $Q(\mathbf{m})$ possess finite moments of all orders. To determine the critical behavior, we expand the self-consistent equations (Eqs.~\ref{self} in the main text) for the forward messages $W^{(+)}$ and $V^{(+)}$ up to the second order around the trivial solution.

Near the percolation threshold, the Taylor expansion yields
\begin{equation}
\begin{aligned}
W^{(+)}& \approx a_WV^{(+)}-\frac{1}{2}b_W(V^{(+)})^2, \\
{V^{(+)}}& \approx a_V W^{(+)}-\frac{1}{2}b_V(W^{(+)})^2,
\end{aligned}
\end{equation}
where the expansion coefficients depend on the moments of the distributions and the anchor node parameter $\theta$
\begin{equation}
\begin{aligned}
a_W&=\frac{\Avg{q^{\rm out}q^{\rm in}}}{\Avg{q^{\rm out}}}, \\
b_W&=\frac{\Avg{q^{\rm out}q^{\rm in}(q^{\rm in}-1)}}{\Avg{q^{\rm out}}}, \\
a_V&=\frac{\Avg{p_H^{[{\bf m}]}p_N{\pi}^{{ m-2}}_N m^{\rm in}m^{\rm out}}}{\Avg{m^{\rm out}}}, \\
b_V&=\frac{\Avg{p_H^{[{\bf m}]}p_N^2{\pi}_N^{{m-3}}m^{\rm out}m^{\rm in}(m^{\rm in}-1)}}{\Avg{m^{\rm out}}}.
\end{aligned}
\end{equation}

Here, $\pi_N = 1-\theta+\theta p_N$ accounts for the survival probability of nodes given the anchor constraint.

By substituting $V^{(+)}$ into the equation for $W^{(+)}$, we obtain a closed-form equation for the order parameter
\begin{equation}
W^{(+)} \approx (a_{W}a_{V}) W^{(+)} - \frac{1}{2}(a_{W}b_{V} + b_{W}a_{V}^{2}) (W^{(+)})^{2}.
\end{equation}
Identifying the branching factor $\hat{\Lambda} = a_{W}a_{V}$ and defining the effective coefficient $C_{W} = a_{W}b_{V} + b_{W}a_{V}^{2}$, this equation simplifies to
\begin{equation}
(\hat{\Lambda} - 1) W^{(+)} - \frac{1}{2} C_{W} (W^{(+)})^{2} \approx 0.
\end{equation}
For the supercritical regime ($0 < \hat{\Lambda} - 1 \ll 1$), the non-trivial solution scales as
\begin{equation}
W^{(+)} \simeq \frac{2}{C_{W}} (\hat{\Lambda} - 1)^{\beta^{(+)}}, \quad \text{with } \beta^{(+)} = 1.
\end{equation}

By following a derivation strictly analogous to that of $W^{(+)}$, one obtains the same critical scaling for $V^{(+)}$. To determine the scaling of the macroscopic order parameters, we expand the expressions for $R^{(+)}$ and $S^{(+)}$ to leading order for small probabilities
\begin{equation}
\begin{aligned}
    R^{(+)}&\approx p_{N} \sum_{q} P(q) q^{\rm in} V^{(+)} = p_{N} \langle q^{\rm in} \rangle V^{(+)}, \\
    S^{(+)}&\approx \sum_{\bf m} p_{H}^{[\bf m]} \pi_{N}^{m} Q({\bf m}) m^{\rm in} W^{(+)}.
\end{aligned}
\end{equation}

Driven by the linear proportionality established above, the macroscopic order parameters $R^{(+)}$ and $S^{(+)}$ inherit the critical scaling of $V^{(+)}$ and $W^{(+)}$, respectively, leading to the same mean-field exponent
\begin{equation}
R^{(+)} \propto (\hat{\Lambda} - 1)^{1}, \quad S^{(+)} \propto (\hat{\Lambda} - 1)^{1}.
\end{equation}

By symmetry, the derivation for the HGIN component involving $W^{(-)}$ and $V^{(-)}$ follows an identical procedure, yielding
\begin{equation}
\beta^{(-)} = 1.
\end{equation}

Finally, to determine the critical exponent for the HGSCC, we expand the self-consistent equations for the order parameters $R$ and $S$ to leading order
\begin{equation}
\begin{aligned}
    R \approx p_{N} \sum_{\mathbf{q}} P(\mathbf{q}) \left[ q^{\rm out} V^{(-)} \right] \left[ q^{\rm in} V^{(+)} \right] = p_{N} \langle q^{\rm in} q^{\rm out} \rangle V^{(+)} V^{(-)}, \\
    S \approx \sum_{\mathbf{m}} p_H^{[\mathbf{m}]} \pi_N^{m} Q(\mathbf{m}) \left[ m^{\rm out} W^{(+)} \right] \left[ m^{\rm in} W^{(-)} \right] \propto W^{(+)} W^{(-)}.
\end{aligned}
\end{equation}

Since the $\langle q^{\rm in} q^{\rm out} \rangle$ and $\langle m^{\rm in} m^{\rm out} \rangle$ are finite in this regime, and given that $V^{(\pm)}$ and $W^{(\pm)}$ share the same critical scaling, both order parameters are governed by the product of the individual scaling behaviors
\begin{equation}
R \sim S \propto (\hat{\Lambda}-1)^{\beta^{(+)}} \cdot (\hat{\Lambda}-1)^{\beta^{(-)}} = (\hat{\Lambda}-1)^{\beta^{(+)} + \beta^{(-)}}.
\end{equation}
This derivation directly validates the additive composition law for the HGSCC critical exponent
\begin{equation}
\beta = \beta^{(+)} + \beta^{(-)}.
\end{equation}
Thus, with $\beta^{(+)}=\beta^{(-)}=1$, we obtain $\beta=2$.

\subsection{Case of uncorrelated degree and cardinality distributions}
\label{uncorr}
We now consider the case where the in/out degrees and cardinalities follow uncorrelated power-law distributions: $P(\bf q) \sim q^{-\gamma_q}$ and $Q(\bf m) \sim m^{-\gamma_m}$. For simplicity, we assume symmetric exponents $\gamma_q^{\rm in} = \gamma_q^{\rm out} = \gamma_q$ and $\gamma_m^{\rm in} = \gamma_m^{\rm out} = \gamma_m$.

The critical behavior is determined by the asymptotic expansion of the self-consistent equations in Eqs.~\ref{self} near the singularity. The form of this expansion depends crucially on whether the second moments of the distributions diverge (i.e., whether $2 < \gamma < 3$).

\subsubsection*{Asymptotic expansions of the self-consistent equations}

First, we examine the equation for $W^{(+)}$. The expansion is governed by the tail of the degree distribution $P(q)$:
\begin{itemize}
    \item For $\gamma_q > 3$, the second moment is finite. The expansion retains the standard linear and quadratic terms
    \begin{equation}
    W^{(+)} \approx a_W V^{(+)} - \frac{1}{2}b_W (V^{(+)})^2.
    \end{equation}
    \item For $2 < \gamma_q < 3$, the second moment diverges. The leading singular term in the expansion scales non-analytically
    \begin{equation}
    W^{(+)} \approx a_W V^{(+)} - d_W (V^{(+)})^{\gamma_q-1},
    \end{equation}
    where $d_W > 0$ is a constant dependent on the distribution tail.
\end{itemize}

Next, we examine the equation for $V^{(+)}$. Here, the presence of anchor nodes plays a critical role in regularizing the cardinality distribution $Q(m)$. The relevant term involves the summation $\sum Q(m) \pi_N^{m-1}$.
\begin{itemize}
    \item If $\theta > 0$, the factor $\pi_N^{m-1}$ (where $\pi_N < 1$) acts as an exponential cutoff, effectively ensuring finite moments for the effective cardinality distribution regardless of $\gamma_m$. Thus, the expansion is always regular
    \begin{equation}
    V^{(+)} \approx a_V W^{(+)} - \frac{1}{2}b_V (W^{(+)})^2.
    \end{equation}
    \item If $\theta = 0$, $\pi_N = 1$. The tail of $Q(m)$ remains purely power-law. Thus, if $2 < \gamma_m < 3$, we observe the singular expansion
    \begin{equation}
    V^{(+)} \approx a_V W^{(+)} - d_V (W^{(+)})^{\gamma_m-1},
    \end{equation}
    where $d_V > 0$ is a constant.
\end{itemize}

To unify these scenarios, we define the effective critical exponent $\gamma$ governing the dominant singularity
\begin{equation}
\gamma = \begin{cases} 
\gamma_q & \text{if } \theta > 0, \\
\min(\gamma_q, \gamma_m) & \text{if } \theta = 0.
\end{cases}
\end{equation}

\subsubsection*{Critical exponents}

By substituting the expansions of $V^{(+)}$ into the equation for $W^{(+)}$ (and vice versa) and solving for the non-trivial solution near $\hat{\Lambda} \approx 1$, we derive the critical exponent $\beta^{(+)}$ defined by $W^{(+)} \propto (\hat{\Lambda}-1)^{\beta^{(+)}}$.

\begin{enumerate}[label=(\roman*), leftmargin=0pt, itemindent=3em]
    \item \textbf{$\gamma > 3$:} The expansions are regular (quadratic).
    \begin{equation}
    \beta^{(+)} = 1.
    \end{equation}
    
    \item \textbf{$2 < \gamma < 3$:} The singular term of order $\gamma - 1$ dominates the quadratic term (since $\gamma - 1 < 2$). The self-consistent equation reduces to the form
    \begin{equation}
    (\hat{\Lambda}-1) W^{(+)} \sim (W^{(+)})^{\gamma-1}.
    \end{equation}
    Solving for $W^{(+)}$, we obtain the anomalous critical exponent
    \begin{equation}
    W^{(+)} \propto (\hat{\Lambda}-1)^{\frac{1}{\gamma-2}} \implies \beta^{(+)} = \frac{1}{\gamma-2}.
    \end{equation}
\end{enumerate}
Due to symmetry, the backward component shares the same exponent, $\beta^{(-)} = \beta^{(+)}$.

Finally, for the HGSCC, the order parameter $R$ scales as the product of the In- and Out-components
\begin{equation}
R \sim S \propto (\hat{\Lambda}-1)^{\beta^{(+)}} (\hat{\Lambda}-1)^{\beta^{(-)}}.
\end{equation}
This yields the additive composition law for the scale-free regime
\begin{equation}
\beta = \beta^{(+)} + \beta^{(-)} = \frac{2}{\gamma-2}.
\end{equation}

This result highlights that for networks with heavy-tailed distributions ($2<\gamma<3$), the transition becomes continuous but with a significantly larger critical exponent compared to the mean-field prediction ($\beta > 2$), implying a smoother transition near the critical point.

\subsection{Case of maximally correlated degree and cardinality distributions}

Finally, we address the case of maximal correlation, where the input and output connectivity of a node (or hyperedge) are strictly coupled. Mathematically, the joint distributions are defined using Dirac delta functions:
\begin{equation}
\begin{aligned}
    P({\bf q})&=\delta({q^{\rm in},q})\delta(q^{\rm out},q)C_q q^{-\gamma_q}, \\
Q({\bf m})&=\delta({m^{\rm in},m})\delta(m^{\rm out},m)C_m m^{-\gamma_m},
\end{aligned}
\end{equation}
where $\gamma_q>2,\gamma_m>2$. For simplicity, we perform the derivation assuming $p^{[{\bf m}]}_H=p_H$ independent of the cardinality ${\bf m}$.

The critical behavior is derived by expanding the self-consistent equations (Eqs.~\ref{self}). Unlike the uncorrelated case, the correlations introduce new singular terms in the expansion.

\subsubsection{Asymptotic expansions}

We first establish the behavior of the $W^{(+)}$ and $V^{(+)}$ for small arguments. The expansions depend on the specific power-law exponents:

\textbf{(1). Expansion for $W^{(+)}$:}
\begin{itemize}
    \item \textbf{$\gamma_q > 4$ (Finite 3rd moment):} The expansion is analytic up to the quadratic term
    \begin{equation}
    W^{(+)} \approx a_W V^{(+)} - \frac{1}{2}b_W (V^{(+)})^2.
    \end{equation}
    \item \textbf{$3 < \gamma_q < 4$ (Diverging 3rd moment):} The quadratic term is overtaken by a singular term
    \begin{equation}
    W^{(+)} \approx a_W V^{(+)} - \hat{d}_W (V^{(+)})^{\gamma_q-2}.
    \end{equation}
    \item \textbf{$2 < \gamma_q < 3$ (Diverging 2nd moment):} The linear term vanishes or becomes sub-dominant to the leading singularity
    \begin{equation}
    W^{(+)} \approx \hat{a}_W (V^{(+)})^{\gamma_q-2}.
    \label{w23}
    \end{equation}
\end{itemize}

\textbf{(2). Expansion $V^{(+)}$:}
Similar to section \ref{uncorr}, the presence of anchor nodes ($\theta > 0$) regularizes the hyperedge distribution.
\begin{itemize}
    \item If $\theta > 0$ (or $\theta=0$ with $\gamma_m > 4$), the expansion remains regular (quadratic)
    \begin{equation}
        V^{(+)} \approx a_V W^{(+)}-\frac{1}{2}b_W[V^{(+)}]^2
        \label{va}
    \end{equation}
    \item If $\theta = 0$ and $3 < \gamma_m < 4$, we obtain the singular expansion
    \begin{equation}
        V^{(+)} \approx a_V W^{(+)} - \hat{d}_V (W^{(+)})^{\gamma_m-2} .
        \label{v34}
    \end{equation}

    \item If $\theta = 0$ and $2 < \gamma_m < 3$, the expansion is dominated by
    \begin{equation}
            V^{(+)} \approx p_H p_N^{\gamma_m-2} \hat{a}_V (W^{(+)})^{\gamma_m-2} .
            \label{v23}
    \end{equation}

\end{itemize}

Here, $\hat{a}_W$, $\hat{d}_W$, $\hat{d}_V$, and $\hat{a}_V$ are constants. For brevity, let $\gamma = \gamma_q$ if $\theta > 0$, and $\gamma = \min(\gamma_q, \gamma_m)$ if $\theta=0$.

In this regime, the critical behavior is governed by the interplay between the exponents $\gamma_q$ and $\gamma_m$. For $\gamma < 3$, the second moments diverge, resulting in heavy-tailed behavior. In this  scenario, the scaling behavior depends on the specific control parameter, necessitating separate analyses for node percolation ($p_H=1$) and hyperedge percolation ($p_N=1$).
\subsubsection{Critical exponents for maximally correlated distributions with $\gamma > 3$}

When $\gamma > 3$, the critical exponents are determined by balancing the leading order terms in the expansion.

\begin{itemize}
    
    \item \textbf{$\gamma > 4$:}
    Both distributions have finite 3rd moment. The expansion is quadratic, yielding the standard result
    \begin{equation}
    \beta^{(+)} = \beta^{(-)} = 1, \quad \beta = 2.
    \end{equation}
    
    \item \textbf{$3 < \gamma < 4$:}
    Here, the singular term with exponent $\gamma-2$ dominates the quadratic term (since $1 < \gamma-2 < 2$). The self-consistent equation behaves as
    \begin{equation}
    (\hat{\Lambda}-1) W^{(+)} \sim (W^{(+)})^{\gamma-2}.
    \end{equation}
    Inverting this relation gives a non-trivial critical exponent
    \begin{equation}
    \beta^{(+)} = \frac{1}{\gamma-3}.
    \end{equation}
    Applying the additive composition law for the HGSCC, we find a very large critical exponent
    \begin{equation}
    \beta = \beta^{(+)} + \beta^{(-)} = \frac{2}{\gamma-3}.
    \end{equation}
\end{itemize}

\subsubsection{Critical exponents for node percolation when $\gamma<3$}
In this regime, the critical behavior is determined by the asymptotic tails of the degree and cardinality distributions. Crucially, the presence of anchor nodes and the intrinsic asymmetry of directed hypergraphs break the duality between node and hyperedge percolation. Consequently, the scaling of the order parameters depends on the specific source of failure, necessitating a separate analysis for node percolation and hyperedge percolation.
For node percolation ($p_H=1$), the order parameters exhibit power-law scalings with respect to the node survival probability $p_N$
\begin{align}
W^{(\pm)} &\propto (p_N)^{\beta_W^{(\pm)}}, & V^{(\pm)} &\propto (p_N)^{\beta_V^{(\pm)}}, \nonumber \\
R^{(\pm)} &\propto (p_N)^{\beta_R^{(\pm)}}, & S^{(\pm)} &\propto (p_N)^{\beta_S^{(\pm)}}, \nonumber \\
R &\propto (p_N)^{\beta_R}, & S &\propto (p_N)^{\beta_S},
\end{align}
where the exponents $\beta_W^{(\pm)}, \beta_{V}^{(\pm)}, \beta_R^{(\pm)}, \beta_{S}^{(\pm)}, \beta_R, \beta_{S}$ are generally different. Specifically, we distinguish the following two scenarios:

\begin{itemize}
\item 
For $\gamma_q < 3$ and $\theta > 0$, or for $\theta=0$ where the divergence is single-sided (either $\gamma=\gamma_m < 3$ and $\gamma_q > 3$, or $\gamma=\gamma_q < 3$ and $\gamma_m > 3$), we have
\begin{equation}
\beta_W^{(\pm)} = \beta_V^{(\pm)} = \frac{\gamma-2}{3-\gamma}.
\end{equation}
These lead to the following critical indices for the order parameters
\begin{align}
\beta_R^{(\pm)} &= 1+\beta_{W}^{(+)}, & \beta_S^{(\pm)} &= \beta_{W}^{(+)}, \nonumber \\
\beta_R &= 1+2\beta_{W}^{(+)}, & \beta_S &= 2\beta_{W}^{(+)}.
\end{align}

\item
For $\theta=0$ and $\gamma < 3$ achieved when both distributions are heavy-tailed ($\gamma_q < 3$ and $\gamma_m < 3$), the interaction of divergences leads to
\begin{align}
\beta_W^{(\pm)} &= \{[(\gamma_q-2)(\gamma_m-2)]^{-1}-1\}^{-1}, \nonumber \\
\beta_V^{(\pm)} &= (\gamma_m-2)(1+\beta_W^{(\pm)}).
\end{align}
These scaling relations determine the critical indices for the macroscopic order parameters
\begin{align}
\beta_R^{(\pm)} &= 1+\beta_{W}^{(+)}, & \beta_S^{(\pm)} &= \beta_{V}^{(+)}, \nonumber \\
\beta_R &= 1+2\beta_{W}^{(+)}, & \beta_S &= 2\beta_{V}^{(+)}.
\end{align}
\end{itemize}

\subsubsection{Critical exponents for hyperedge percolation when $\gamma<3$}

Similarly, for hyperedge percolation ($p_N=1$) in the $\gamma < 3$ regime, we observe the scalings with respect to the hyperedge survival probability $p_H$
\begin{align}
W^{(\pm)} &\propto (p_H)^{\beta_W^{(\pm)}}, & V^{(\pm)} &\propto (p_H)^{\beta_V^{(\pm)}}, \nonumber \\
R^{(\pm)} &\propto (p_H)^{\beta_R^{(\pm)}}, & S^{(\pm)} &\propto (p_H)^{\beta_S^{(\pm)}}, \nonumber \\
R &\propto (p_H)^{\beta_R}, & S &\propto (p_H)^{\beta_S}.
\end{align}
Depending on the topological conditions, we have the following scenarios:

\begin{itemize}
\item
For $\gamma_q < 3$ and $\theta > 0$, or for $\theta=0$ with a single-sided divergence (as defined above), the exponents are
\begin{equation}
\beta_W^{(\pm)} = \beta_V^{(\pm)} = 1/(3-\gamma).
\end{equation}
These lead to the following critical indices for the order parameters
\begin{align}
\beta_R^{(\pm)} &= \beta_{W}^{(+)}, & \beta_S^{(\pm)} &= 1+\beta_{W}^{(+)}, \nonumber \\
\beta_R &= 2\beta_{W}^{(+)}, & \beta_S &= 1+2\beta_{W}^{(+)}.
\end{align}

\item
For $\theta=0$ and $\gamma < 3$ driven by double divergence ($\gamma_q < 3$ and $\gamma_m < 3$), we obtain
\begin{align}
\beta_V^{(\pm)} &= [1-(\gamma_q-2)(\gamma_m-2)]^{-1}, \nonumber \\
\beta_W^{(\pm)} &= (\gamma_q-2)\beta_V^{(\pm)}.
\end{align}
This results in the following macroscopic critical indices
\begin{align}
\beta_R^{(\pm)} &= \beta_{W}^{(+)}, & \beta_S^{(\pm)} &= 1+\beta_{V}^{(+)}, \nonumber \\
\beta_R &= 2\beta_{W}^{(+)}, & \beta_S &= 1+2\beta_{V}^{(+)}.
\end{align}
\end{itemize}

It is important to emphasize that the fundamental additive scaling law for the strongly connected component ($\beta = \beta^{(+)} + \beta^{(-)}$) requires a specific correction when the order parameter explicitly depends on the driving control parameter.

\section{Numerical Validation of Critical Exponents}
\label{ce}
While the Finite-Size Scaling analysis presented in Fig.~\ref{fig:FSS} provides rigorous validation for systems with finite moments, the verification of large critical exponents presents specific computational challenges. Theoretical analysis predicts that under specific topological conditions, the critical exponents $\beta^{(\pm)}$ and $\beta$ can become relatively  large($\beta > 1$). In these cases, the order parameter grows extremely slowly near the critical threshold, meaning that the asymptotic scaling region is narrow and heavily affected by finite-size effects. Consequently, a traditional FSS collapse would require simulating prohibitively large system sizes to isolate the critical behavior from noise.

To validate our theoretical predictions in these anomalous regimes, we adopted a direct asymptotic analysis approach. We performed Monte Carlo simulations on large-scale directed hypergraphs ($N=10^5$) to approximate the thermodynamic limit. Instead of scaling the pseudo-critical point, we fixed the control parameter $p_N$ and measured the equilibrium sizes of the giant components $R^{(\pm)}$ and $R$ in the close vicinity of the theoretical percolation threshold $p_c$.

By plotting the order parameters against the distance from the critical point on a logarithmic scale, $\ln(R)$ vs. $\ln(p_N - p_c)$, we extracted the critical exponents from the slope of the linear fit. The results, presented in Figs. S1 to S5, confirm the validity of our analytical framework across different topological classe.

\begin{figure}
    \centering
    \includegraphics[width=1\linewidth]{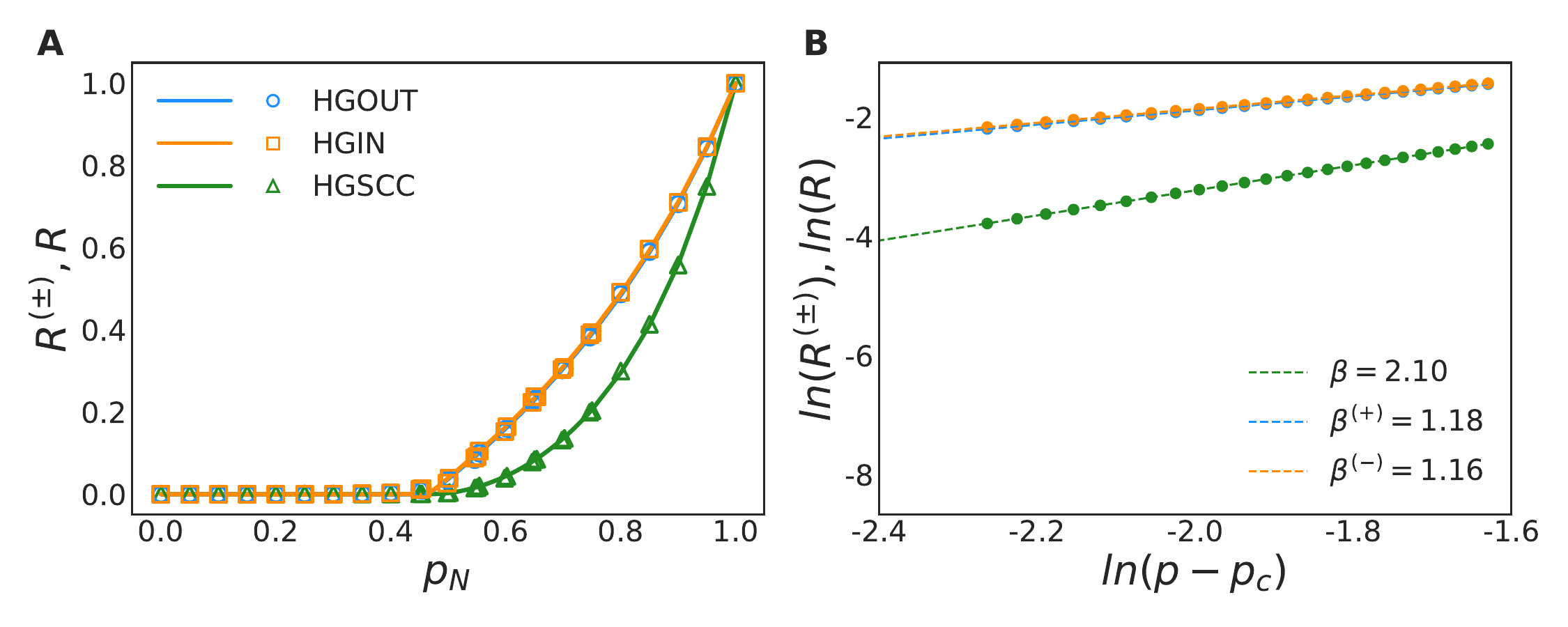}
    \caption{\textbf{Numerical validation of critical exponents for uncorrelated directed hypergraphs with $\gamma=3.5$.}
(A) The macroscopic order parameters $R^{(+)}$, $R^{(-)}$, and $R$ as a function of the node survival probability $p_N$. (B) Log-log plot of the order parameters versus the distance from the critical threshold, $\ln(p_N - p_c)$. The symbols represent Monte Carlo simulation data obtained from synthetic hypergraphs. The dashed lines indicate the linear fits in the critical scaling region. The extracted exponents $\beta^{(\pm)} \approx 1.0$ and $\beta \approx 2.0$ confirm the mean-field predictions for distributions with finite second moments.}
    \label{un3.5}
\end{figure}

\textbf{Uncorrelated Distributions (Finite Moments with $\gamma > 3$):} For uncorrelated power-law distributions with $\gamma = 3.5$ (Fig. \ref{un3.5}), where the exponents take standard values, we recover the mean-field prediction. The numerical results yield $\beta^{(\pm)} \approx 1.0$ and $\beta \approx 2.0$, perfectly matching the theoretical prediction $\beta = \beta^{(+)} + \beta^{(-)} = 2$.

\begin{figure}
    \centering
    \includegraphics[width=1\linewidth]{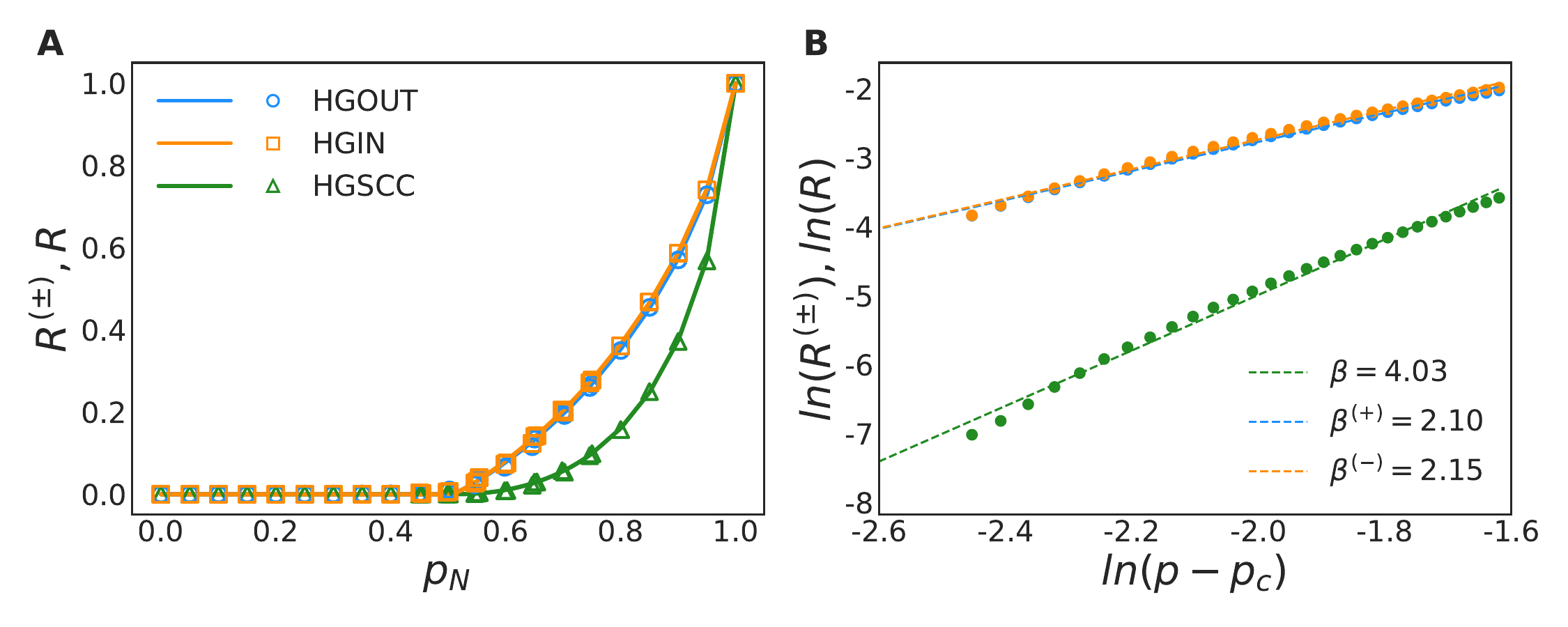}
    \caption{\textbf{Numerical validation of anomalous critical exponents for uncorrelated directed hypergraphs with $\gamma=2.5$.}
(A) Equilibrium sizes of the giant components as a function of $p_N$. (B) Asymptotic scaling analysis on a logarithmic scale. The system is characterized by power-law degree and cardinality distributions with exponents $\gamma_q = \gamma_m = 2.5$. The numerical fits yield $\beta^{(\pm)} \approx 2.1$ and $\beta \approx 4.03$.}
    \label{un2.5}
\end{figure}

\textbf{Uncorrelated Distributions ($2 < \gamma < 3$)}:
For $\gamma = 2.5$ (Fig. \ref{un2.5}), the theory predicts larger anomalous exponents. Our derivation  predicts $\beta^{(\pm)} = 1/(\gamma-2) = 2.0$, implying $\beta = 4.0$. The numerical fits reveal $\beta^{(\pm)} \approx 2.1$ and $\beta \approx 4.03$.

\begin{figure}
    \centering
    \includegraphics[width=1\linewidth]{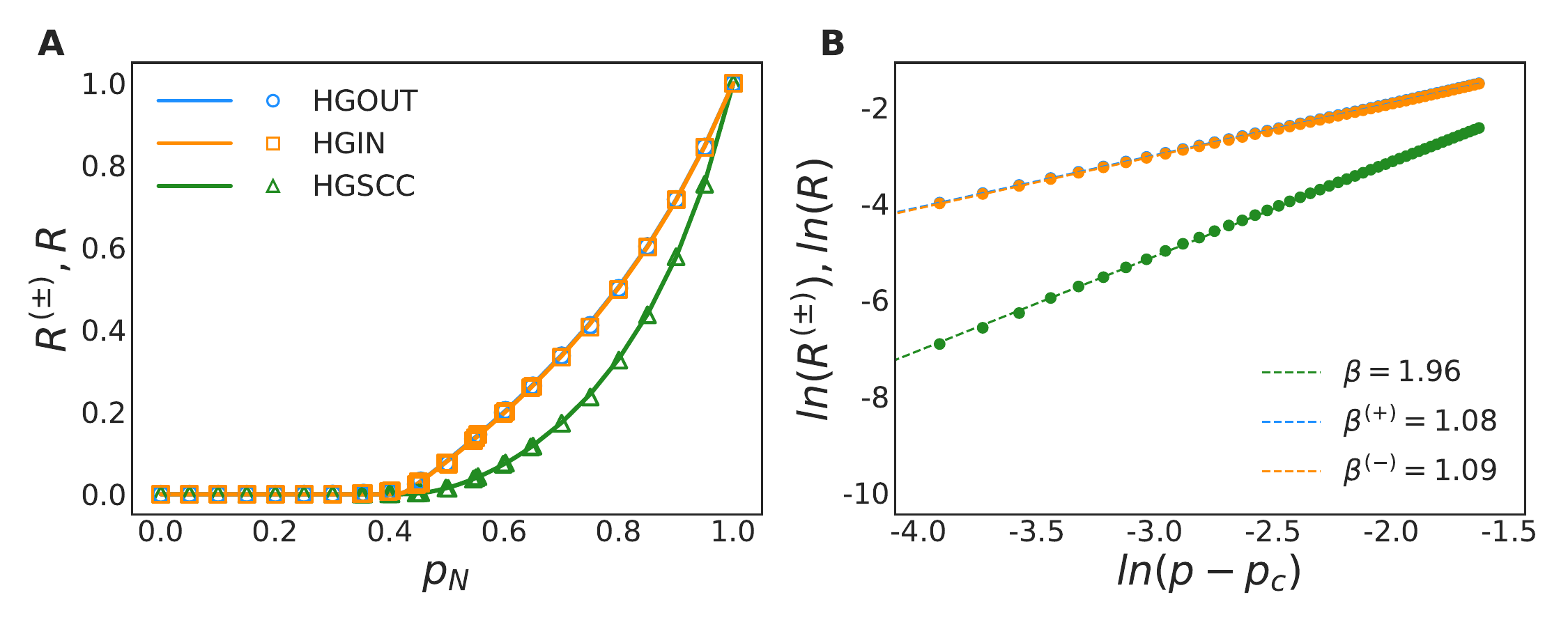}
    \caption{\textbf{Numerical validation of anomalous critical exponents for maximally correlated directed hypergraphs with $\gamma=4.5$.}
(A) Percolation phase transition of the HGOUT, HGIN, and HGSCC. (B) Log-log scaling of the order parameters near the critical point. In this regime ($\gamma > 4$), the correlations do not alter the universality class. The simulation results yield $\beta^{(\pm)} \approx 1.0$ and $\beta \approx 1.96$.}
    \label{max4.5}
\end{figure}

\textbf{Maximally Correlated Distributions ($\gamma > 4$):}
In the correlated regime with $\gamma = 4.5$ (Fig. \ref{max4.5}), the system exhibits standard scaling. The simulations confirm $\beta^{(\pm)} \approx 1.0$ and $\beta \approx 1.96$.

\begin{figure}
    \centering
    \includegraphics[width=1\linewidth]{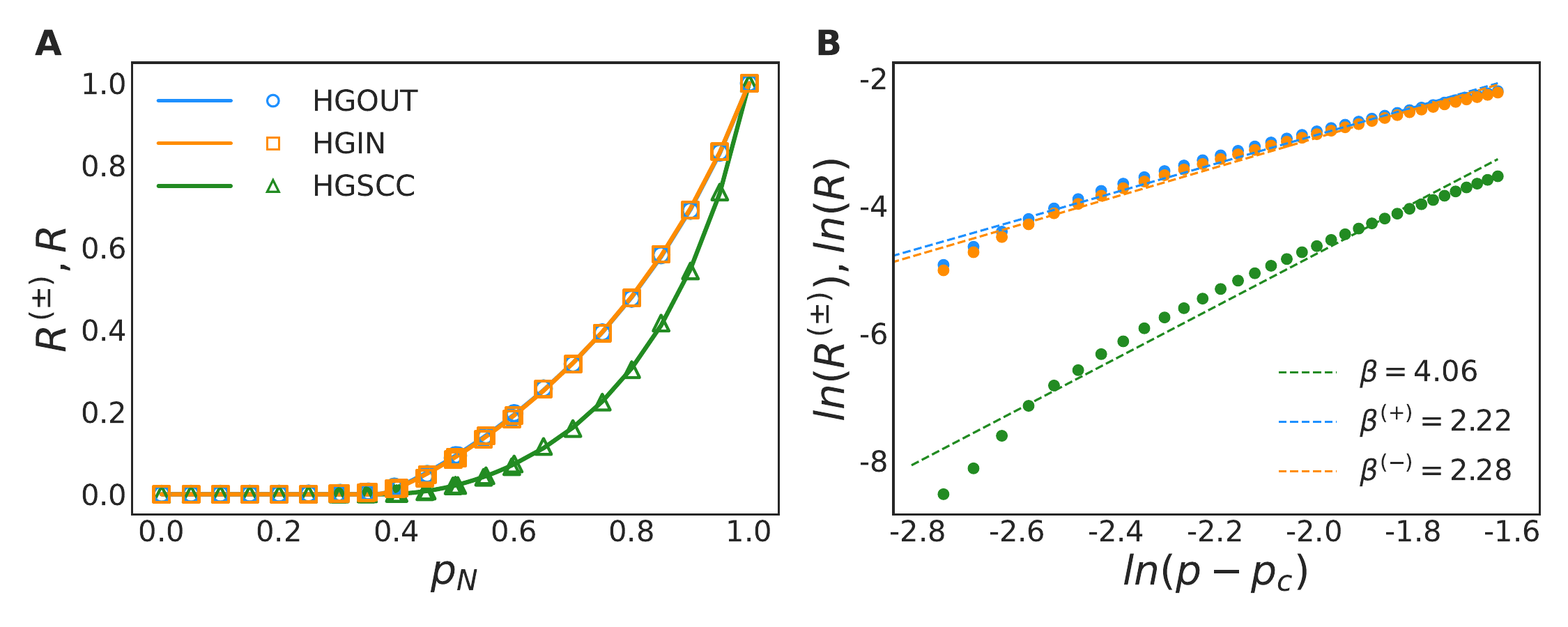}
    \caption{\textbf{Numerical validation of anomalous critical exponents for maximally correlated directed hypergraphs with $\gamma=3.5$.}
(A) The emergence of giant components relative to the control parameter $p_N$. (B) Determination of critical exponents via linear fitting on a double-logarithmic scale. The theoretical predictions for this regime are $\beta^{(\pm)} = 1/(\gamma-3) = 2.0$ and $\beta = 2/(\gamma-3) = 4.0$. The linear fits to the simulation data yield the measured values $\beta^{(\pm)} \approx 2.2$ and $\beta \approx 4.06$.}
    \label{max3.5}
\end{figure}

\textbf{Maximally Correlated Distributions ($3 < \gamma < 4$):}
For $\gamma = 3.5$ (Fig. \ref{max3.5}), the exponents increase. Our theory predicts $\beta^{(\pm)} = 1/(\gamma-3) = 2.0$ and $\beta = 4.0$. The simulation results closely track these predictions with $\beta^{(\pm)} \approx 2.2$ and $\beta \approx 4.06$, validating the generalized scaling in correlated structures.

\begin{figure}
    \centering
    \includegraphics[width=1\linewidth]{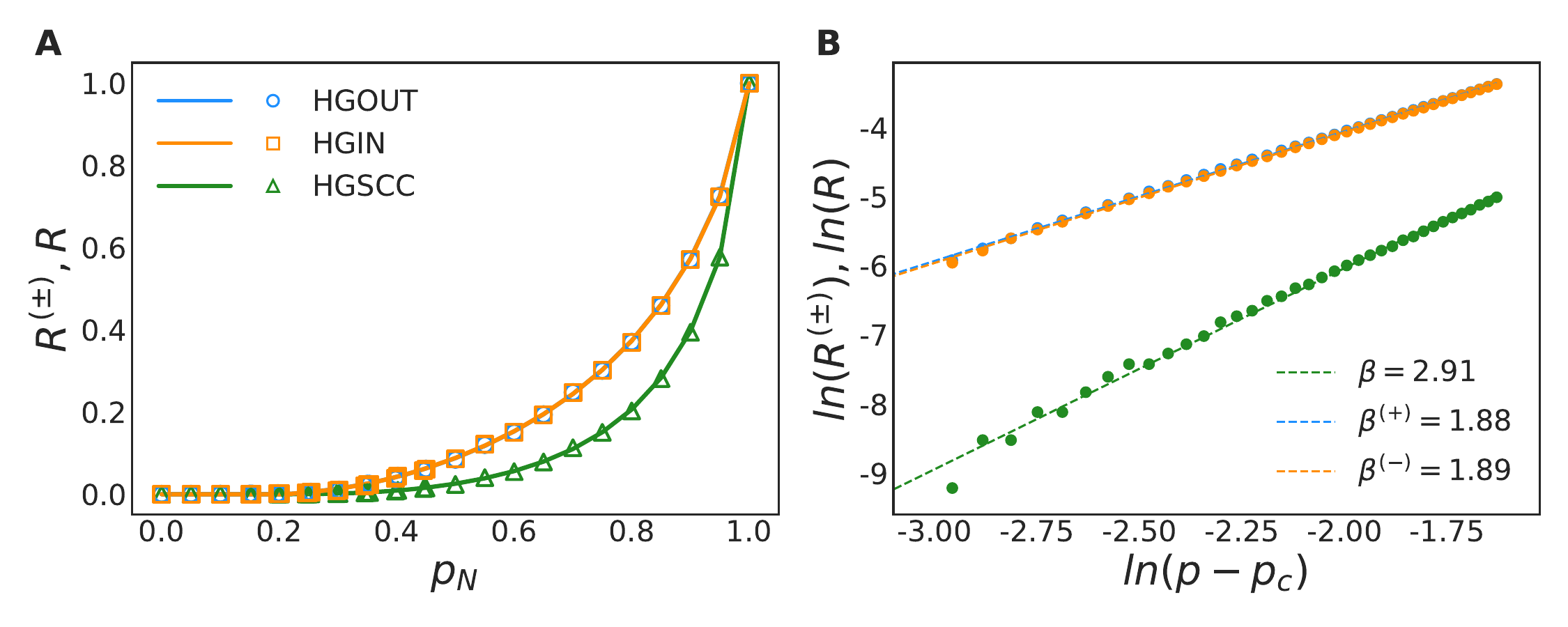}
    \caption{\textbf{Numerical validation of critical exponents in maximally correlated directed hypergraphs ($\gamma=2.5$, Node Percolation).}
(A) Order parameters vs. node survival probability $p_N$. (B) Log-log scaling of the order parameters near the critical point. The data corresponds to node percolation on maximally correlated hypergraphs with $\gamma=2.5$. The fitted slopes yield $\beta_R^{(\pm)} \approx 1.88$ and $\beta_R \approx 2.91$. These results confirm the modified scaling relation $\beta_R = \beta_R^{(+)} + \beta_R^{(-)} - 1$.}
    \label{max2.5}
\end{figure}

\textbf{Maximally Correlated Distributions ($\gamma < 3$, Node Percolation):}
For $\gamma = 2.5$ (Fig. \ref{max2.5}), where the  universality breaks. Based on the analysis, we expect $\beta_R^{(\pm)} \approx 2.0$ and $\beta_R \approx 3.0$. The numerical data yields $\beta_R^{(\pm)} \approx 1.88$ and $\beta_R \approx 2.91$. The slight deviation is attributed to strong finite-size corrections inherent to these topologies; however, the results clearly uphold the modified scaling relation $\beta_R = \beta_R^{(+)} + \beta_R^{(-)} - 1$.

In summary, these direct asymptotic measurements confirm that the message-passing theory accurately captures the critical behavior of directed hypergraphs, validating both the additive composition law ($\beta = \beta^{(+)} + \beta^{(-)}$) and its corrections.
\newpage


\begin{figure}
    \centering
    \includegraphics[width=0.6\linewidth]{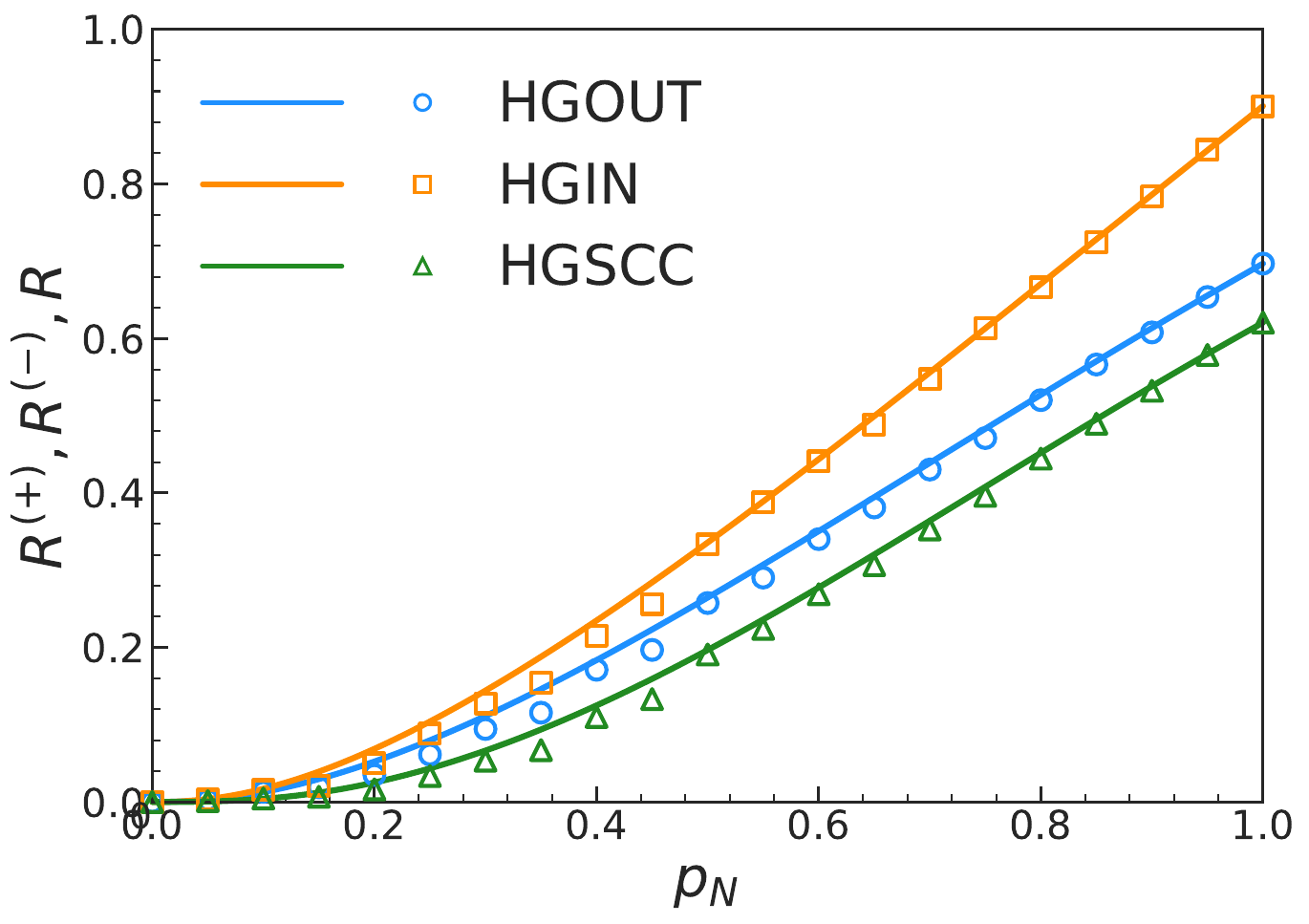}
    \caption{\textbf{Percolation analysis of the E. coli metabolic hypergraph in the anchor-free limit.}
The relative sizes of the HGOUT ($R^{(+)}$, blue circles), HGIN ($R^{(-)}$, orange squares), and HGSCC ($R$, green triangles) are plotted as a function of the node survival probability $p_N$. Solid lines correspond to the analytical predictions derived from the message-passing theory, while symbols represent Monte Carlo simulations. This null model assumes no functional dependencies, illustrating the purely structural robustness of the metabolic network topology in contrast to the biologically constrained case shown in Fig. \ref{fig:ijo}.}
    \label{fig:nonanchor}
\end{figure}
\end{document}